\documentclass[12pt,preprint]{aastex}
\slugcomment{Submitted to The Astrophysical Journal, Part 1.}

\shorttitle{3D MHD Supernova}
\shortauthors{Mikami et al.}

\begin{document}

\title{THREE DIMENSIONAL MAGNETOHYDRODYNAMICAL SIMULATIONS OF
CORE COLLAPSE SUPERNOVA}

\author{Hayato Mikami and Yuji Sato}
\affil{Graduate School of Science, Chiba University,
1-33, Yayoi-cho, Inage-ku, Chiba 263-8522, Japan}
\email{mikami@astro.s.chiba-u.ac.jp}

\author{Tomoaki Matsumoto}
\affil{Faculty of Humanity and Environment,
Hosei University, Fujimi, Chiyoda-ku,
Tokyo 102-8160, Japan}

\and

\author{Tomoyuki Hanawa}
\affil{Center for Frontier Science, Chiba University,
1-33, Yayoi-cho, Inage-ku, Chiba 263-8522, Japan}

\begin{abstract}
We show three-dimensional magnetohydrodynamical simulations of core collapse supernova in which the progenitor has magnetic fields inclined to the rotation axis.  The simulations employed a simple empirical equation of state in which the pressure of degenerate gas is approximated by piecewise polytropes for simplicity.  Neither energy loss due to neutrino is taken into account for simplicity.  The simulations start from the stage of dynamical collapse of an iron core.  The dynamical collapse halts at $ t $~=~189~ms by the pressure of high density gas and a proto-neutron star (PNS) forms.  The evolution of PNS was followed about 40~milli-seconds in typical models.  When the initial rotation is mildly fast and the initial magnetic fields are mildly strong, bipolar jets are launched from an upper atmosphere ($ r \, \sim \, 60~{\rm km} $) of the PNS.  The jets are accelerated to $ \sim 3 \times 10 ^4 $~km~s$^{-1}$, which is comparable to the escape velocity at the foot point.  The jets are parallel to the initial rotation axis.  Before the launch of the jets, magnetic fields are twisted by rotation of the PNS.  The twisted magnetic fields form torus-shape multi-layers in which the azimuthal component changes alternately.  The formation of magnetic multi-layers is due to the initial condition in which the magnetic fields are inclined with respect to the rotation axis.  The energy of the jet depends only weakly on the initial magnetic field assumed.  When the initial magnetic fields are weaker, the time lag is longer between the PNS formation and jet ejection.  It is also shown that the time lag is related to the Alfv\'en transit time.  Although the nearly spherical prompt shock propagates outward in our simulations, it is an artifact due to our simplified equation of state and neglect of neutrino loss.  The morphology of twisted magnetic field and associate jet ejection are, however, not affected by the simplification.
\end{abstract}

\keywords{accretion, accretion disks  --- methods: numerical ---
MHD --- supernovae: general}

\section{INTRODUCTION}

Explosion mechanism of core collapse supernova has been an
open question for more than three decades.  Numerical
simulations have not succeeded in constructing a convincing
model of core collapse supernova, although they have been
updated and improved steadily.  The problem is likely
to be neither a simple numerical error nor inaccuracy of
neutrino transfer.  It has been  thought that
multi-dimensional effects play essential roles in the
explosion \citep[see, e.g.,][and the references therein]{burrows07}.
This is based on the fact that
the spherical symmetric model cannot reproduce explosion
while it has been sophisticated to an extreme.  At the same
time, observational evidences have been accumulated for
inherent non-spherical natures of core collapse supernova
\citep[see, e.g,][]{wang02,wang03,hwang04,leonard06}.

Magnetic fields and rotation have been thought to be an
agent to promote global non-sphericity, although
it can be produced without magnetic fields
through some other mechanisms proposed by
\citet{burrows06}, \citet{blondin07}, and others.
As shown by earlier numerical simulations, magnetic fields twisted by
rotation produces high velocity bipolar jets, if the initial magnetic
field is relatively strong and initial rotation is fast.
Since \citet{leblanc70}, many magnetohydrodynamical
simulations of core collapse supernovae have been published
\citep*[see, e.g,][]
{yamada04,ardeljan04,takiwaki04,sawai05,moiseenko06,shibata06,
obergaulinger06a,obergaulinger06b,burrows07}.
However all of them are two-dimensional and
have assumed symmetry around the axis.  The symmetry excludes
the possibility that magnetic fields are inclined to the
rotation axis, although pulsars are believed to have such
magnetic fields.

In this paper we show three dimensional numerical simulations
of core collapse supernova.  The initial magnetic field is assumed
to be inclined with respect to the rotation axis.  It is
assumed to be stronger than expected from a standard evolutionary model
in part because a weak magnetic field can have dynamical effects only long
afterward and in part because it can be strong enough in some
circumstances.  Such a strong magnetic field may be realized
in progenitors of magnetars.
In a typical model of our simulations, a magnetic torus is formed
around the PNS and magnetohydrodynamical jets are launched
along the initial rotation axis.  It is also shown that
the toroidal component of the magnetic field
changes its sign alternately in the magnetic torus.
We also discuss the dependence on the initial magnetic field
strength, the initial angular velocity, and initial inclination
angle.  When the initial magnetic fields are weaker, the jets
are launched at a later epoch.  The total energy of the jets
depends only weakly on the initial magnetic fields.

In \S 2 we summarize our basic model and numerical methods.
The results of numerical simulations are shown in \S 3.
Discussions are given in \S 4.  Appendix is devoted to the
numerical schemes that we have developed for the numerical
simulations.

\section{MODEL AND METHODS OF COMPUTATION}

\subsection{Basic Equations}

As a model of core collapse supernova, we consider gravitational
collapse of a massive star with taking account of magnetic field.
The dynamics is described by the Newtonian ideal magnetohydrodynamical
(MHD) equations,
\begin{equation}
\frac{\partial \rho}{\partial t}  + \mbox{\boldmath$\nabla$} \cdot
(\rho \mbox{\boldmath$v$}) = 0 ,
\end{equation}
\begin{equation}
\frac{\partial \mbox{\boldmath$v$}}{\partial t} +
(\mbox{\boldmath$v$}\cdot\mbox{\boldmath$\nabla$})
\mbox{\boldmath$v$} +
\frac{1}{\rho} \left[ \mbox{\boldmath$\nabla$} P
- \left(\frac{\mbox{\boldmath$\nabla$}\times\mbox{\boldmath$B$}}
{4\pi} \right) \times \mbox{\boldmath$B$} \right]
 - \mbox{\boldmath$g$} = 0 ,
\end{equation}
\begin{equation}
\frac{\partial \mbox{\boldmath$B$}}{\partial t} =
\mbox{\boldmath$\nabla$} \times \left(
\mbox{\boldmath$v$} \times \mbox{\boldmath$B$}
\right) ,
\end{equation}
and
\begin{equation}
\mbox{\boldmath$g$} = - \mbox{\boldmath$\nabla$} \Phi
,
\end{equation}
where $ \rho $, $P$, $ \mbox{\boldmath$v$}$,
$ \mbox{\boldmath$B$}$, $\mbox{\boldmath$g$}$, and $ \Phi $ denote
the density, pressure, velocity, magnetic field, gravity, and gravitational
potential, respectively.  The gravitational potential, $\Phi$, is given by the
Poisson equation,
\begin{equation}
\Delta \Phi = 4 \pi G \rho .
\end{equation}

We used the equation of state of \citet{takahara82} in which
the pressure is expressed as
\begin{eqnarray}
P & = & P _{\rm c} \, + \, P _{\rm t} \, , \label{eos1}\\
P _{\rm t} & = &
\frac{\rho \varepsilon _{\rm t}}{\gamma _{\rm t} \, - \, 1} \, ,
\label{eos2}
\end{eqnarray}
and
\begin{equation}
P _{\rm c}  = K _i \, \left(
\frac{\rho}{\rho _i} \right) ^{\gamma _i}\, . \label{eos3}
\end{equation}
The index, $ \gamma _t $, is taken to be 1.3.
The coefficients, $ K _i $ and $ \gamma _i $, are
piecewise constant in the interval of
$ \rho _{i-1} \, < \, \rho \, \le \, \rho _i $.
The values are given in Table \ref{eos-table}.
The internal energy per unit mass is expressed as
\begin{equation}
\varepsilon \; = \; \varepsilon _t \, + \,
\int _0 ^\rho \frac{P _{\rm c}}{\rho ^2} \, d\rho \, .
\end{equation}

Accordingly we have the equation of energy conservation,
\begin{equation}
\frac{\partial}{\partial t} (\rho E) + \mbox{\boldmath$\nabla$}
\cdot \left( \rho H \mbox{\boldmath$v$} \right)\,
= \rho \mbox{\boldmath$v$}\cdot\mbox{\boldmath$g$} \, ,
\label{energy-conservation}
\end{equation}
where the specific energy ($ E $) and specific enthalpy ($ H $)
are expressed as,
\begin{equation}
E = \frac{\mbox{\boldmath$v$} ^2}{2} +
\int _0 ^\rho \frac{P _c}{\rho ^2} d\rho +
\varepsilon _t + \frac{\mbox{\boldmath$B$} ^2}{8\pi} ,
\end{equation}
and
\begin{equation}
H = E + \frac{P}{\rho} + \frac{\mbox{\boldmath$B$} ^2}{8\pi} ,
\end{equation}
respectively.

\subsection{Numerical Grid}

We solved the MHD equations and the Poisson equation
simultaneously on a nested grid.  The nested grid
covers a rectangular box of ($ 3.39 \times 10 ^3 $~km)$^3$
with resolution of $ \Delta x \, = \, 52.9 \, {\rm km}$.
The central eighth volume is covered hierarchically with a finer
grid of which cell width is the half of the coarse grid.
We overlapped rectangular grids of 8 different resolutions
and achieved very high resolution of
$ \Delta x \, = \, 0.413 \, {\rm km} $ for the central cube
of $ (26.0 \, {\rm km}) ^3 $.  The finest grid fully covers the
PNS.  Since the nested grid has $64^3$ cells
at each level, the resolution is roughly proportional to the
radius from the center and approximately 4 \% of the
radius.  This angular resolution is comparable to that of
recent two-dimensional simulations,
since the angular resolution is $ \Delta \theta \, = \,
(\pi/2) / 30 \, = \, 5.24 \times 10 ^{-2} $ in most of them
and $ \Delta \theta \, = \, (\pi/2) /71 \, = \, 2.21\times 10^{-2}$
in \citet{burrows07}.
We call this hierarchically arranged grids
the nested grid.
All the physical quantities are evaluated at the cell center
except for the magnetic field.
The divergence-free staggered mesh of \citet{balsara01} and \citet{balsara99}
is employed for the magnetic field in order to keep
$\mbox{\boldmath$\nabla$} \cdot \mbox{\boldmath$B$} = 0$
within a round-off error.
This method is a variant of the constrained transport approach of
\citet{evans88}, and is optimized for the Godunov-type Riemann solver
and hierarchical grids.
The same type of nested grid has been used
for formation of protostars from a cloud core \citep{matsumoto04}.

The outer boundary condition is set at the sphere of
which radius is $ r \, = \, 1.66 \times 10 ^3 $~km.
The density, pressure, velocity, and magnetic fields are
fixed at the initial values outside the boundary.

\subsection{Numerical Scheme}

A \citet{roe81}-type approximate Riemann solution is
employed to solve the MHD equations.  It takes account of
the cold pressure, $ P _c $.  Thus it is slightly
different from that of \citet{cargo97} which is
designed to satisfy the property U of \citet{roe81} for
an ideal gas.  The details are given in Appendix.

We adopted supplementary numerical viscosity to care
the carbuncle instability since the Roe-type scheme
is vulnerable.  The supplementary viscosity has a large
value only near shocks.  The detailed form of the viscosity
is given in \citet{hanawa07}.

The source term in the equation of energy conservation,
$ \rho \mbox{\boldmath$v$} \cdot \mbox{\boldmath$g$} $, is
evaluated to be the inner product of the gravity
and the average numerical mass flux.  In other words, we
evaluated the mass flux, $ \rho \mbox{\boldmath$v$} $, not at the
cell center but on the cell surface.  By the virtue of this evaluation,
the source term vanishes when the mass flux vanishes.
Note that the mass flux evaluated at the cell center may not
vanish in a Roe type scheme even when that evaluated on the cell
surface vanishes.  We have found that PNS
suffers from serious spurious heating when the source term
is evaluated from the cell center density and velocity.
The spurious heating expands PNS to blow off eventually.

The Poisson equation was solved by the nested grid iteration
\citep{matsumoto03} as in the simulations of protostar formation.

\subsection{Initial Model}

Our initial model was constructed from the 15~M$_\odot$ model of
\citet{woosley02}.  The initial density is increased
10~\% artificially to initiate the dynamical collapse.
Thus it is $\rho_0 = 6.8 \times 10^9 {\rm ~g~cm^{-3}}$
at the center.
Their model assumes the spherical symmetry and
takes account of neither rotation nor magnetic field.
We have constructed our initial model by adding a dipole
magnetic field and nearly solid rotation.

The initial magnetic field is assumed to be
\begin{equation}
\left(\begin{array}{c}
B _r \\
B _\theta \\
B _\varphi
\end{array}\right)
=
B_0
\left(
\begin{array}{c}
\cos \theta \\ -\sin \theta \\ 0
\end{array} \right)
\end{equation}
in the central core of $ r \, \le \, r _a $,
\begin{equation}
\left( \begin{array}{c}
B _r \\
B _\theta \\
B _\varphi
\end{array}
\right) =
\frac{B _0}{8}
\left[
\begin{array}{c}
\displaystyle
\left( 16 - \frac{6 r}{r_a} - \frac{2 r_a^3}{r ^3} \right)
\, \cos \theta \\
\displaystyle
- \left( 16  -  \frac{9 r}{r_a}  + \frac{r_a^3}{r^3} \right)
\sin \theta \\ 0
\end{array} \right] \, ,
\end{equation}
in the middle region of $ r _a \, \le \, r \, \le \, 2 \, r _a $,
and
\begin{equation}
\left( \begin{array}{c}
B _r \\
B _\theta \\
B _\varphi
\end{array}
\right) =
 \frac{15 \, B_0  \, r _a ^3 }{8 \, r^3}
\left(
\begin{array}{c}
2 \cos \theta \\  \sin \theta \\ 0
\end{array} \right)
\end{equation}
in the outer region of $ 2 \, r _a \, \le \, r $,
in the spherical coordinates,
where $ r_a = 846$ km.  Thus the initial magnetic field is
uniform inside $ r \, \le \, r _a $, while it is dipolar
outside $ r \, \ge \, r _a $.  The uniform and dipole fields
are connected without kink so that the magnetic tension force
is finite.
The electric current density is uniform
in the transition region of $ r _a \le \, r \, < \, 2 \, r _a$
in this magnetic configuration.

The initial rotation velocity is expressed to be
\begin{equation}
\mbox{\boldmath$v$} _0 \; = \; \Omega (r) \,
(\mbox{\boldmath$e$} _\Omega \times
\mbox{\boldmath$r$}) \, ,
\end{equation}
where
\begin{equation}
\Omega \,(r) \; = \;
\frac{\Omega _0 \, a ^2}{r ^2 \, + \, a ^2} \, ,
\end{equation}
\begin{equation}
\mbox{\boldmath$e$} _\Omega \; = \;
\left( \begin{array}{c}
0
\\
- \sin \, \theta _\Omega
\\
\cos \, \theta _\Omega
\end{array} \right) \, ,
\end{equation}
and
\begin{equation}
\mbox{\boldmath$r$} \; = \;
\left( \begin{array}{c}
x \\ y \\ z
\end{array} \right) \, ,
\end{equation}
in the Cartesian coordinates.  The rotation axis is
inclined by $ \theta _\Omega $ from the $ z $-axis,
i.e., from the magnetic axis.

The initial central magnetic field is set in the range of
$ 1.7 \times 10 ^{11} $~G $ \le B _0 \le \, 2.0 \times 10^{12} $~G
except in model R0B0.
The initial angular velocity is set in the range of
$ 0.31~{\rm s}^{-1} \, \le \, \Omega _0 \, \le \, 1.21~{\rm s}^{-1} $
except in model R0B0.  The models computed are summarized in
Table~\ref{modeltable}.

The assumed initial magnetic field is stronger than those
evaluated by \citet{heger05}.  Our choice is based
on the constraint that our three-dimensional numerical
simulations can follow only several tens milliseconds after
the bounce.  When the initial magnetic field is weak, it
cannot have any dynamical effects on a relatively short
timescale even if it is amplified through collapse and
rotation.  Thus we have assumed rather strong initial
magnetic field, which can be realized in some progenitors.

\section{RESULTS}

\subsection{Non-rotating Non-magnetized Model}

First we show model R0B0 having no magnetic field and no rotation
as a test of our numerical code.  In this model the central
density increases to reach the maximum value,
$\rho _{\max} \, = \, 5.71 \times 10 ^{14} $~g~cm$^{-3}$, at
$ t $~=~187.11~ms.  It settles down to the equilibrium
value, $\rho _{\rm eq} \, = \, 4.66 \times 10 ^{14} $~g~cm$^{-3}$,
after several times of radial oscillation.  The oscillation
period is 1.25~ms.  Radial shock waves are initiated by this
core bounce.  The first one, the prompt shock wave, reaches
$ r \, = \, 595$~km at $ t $~=~222.71~ms, where
$ r $ denotes the radial distance from the center.
It reaches the boundary of computation
($ r \, = \, 1700~{\rm km}$) at $ t \, = \, 303~{\rm ms}$,
while the expansion velocity decreases.
The fast propagation of prompt shock is an artifact due to
our simplified EOS.  If we had taken neutrino transfer into
account, the prompt shock should have stalled around
$ r \, \simeq \, 100~{\rm km} $ roughly within 100~ms
after the bounce \citep[see, e.g.][]{sumiyoshi05,buras06,burrows06}.

\begin{figure}
\begin{center}
\includegraphics[width=0.5\textwidth]{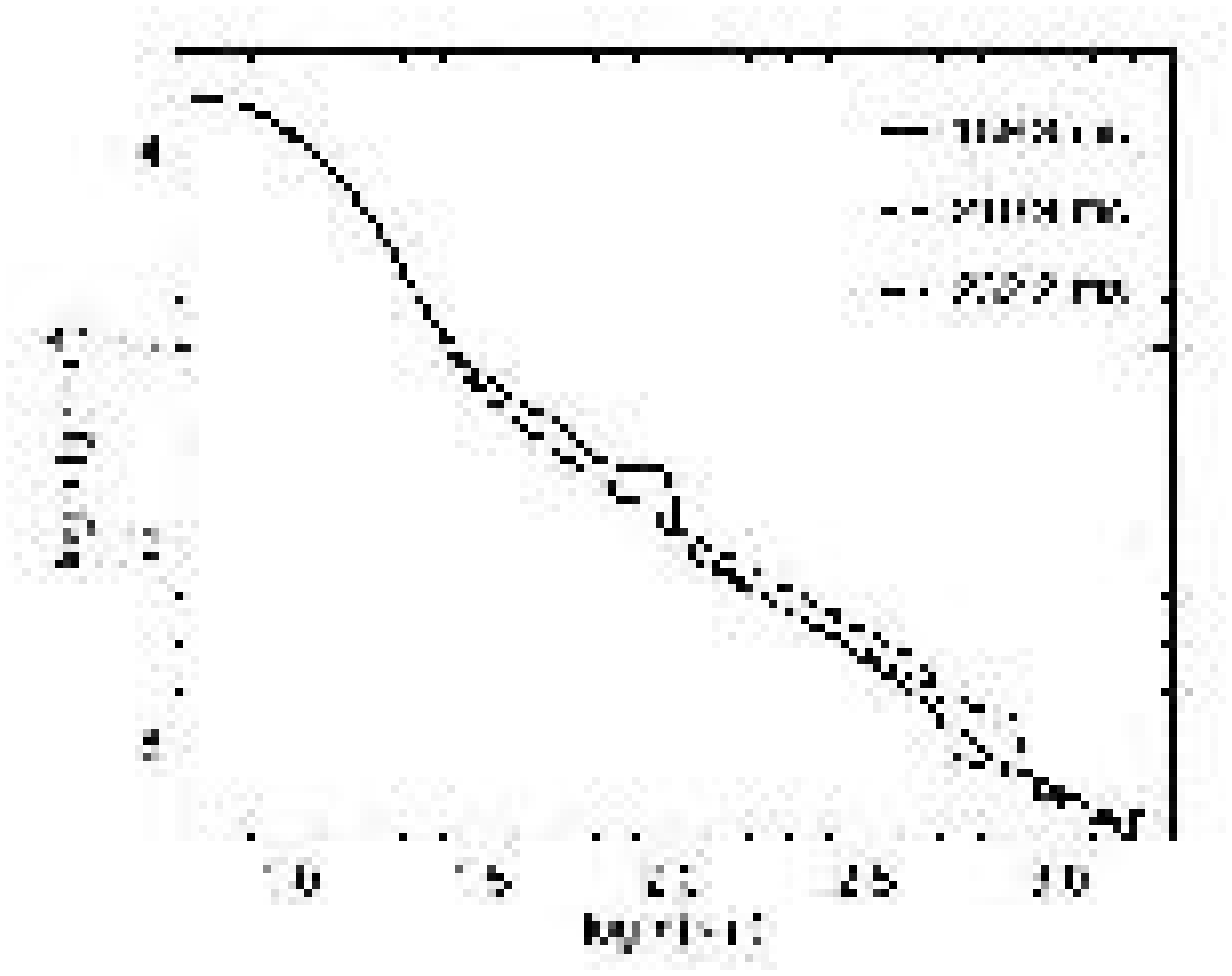}
\end{center}
\caption{The angle averaged density, $ \rho \,(r,~t)$, is shown
as a function of $ r $ for model R0B0.
\label{rho-rt-r0b0}}
\end{figure}

Figure~\ref{rho-rt-r0b0} shows the radial density profiles
at $ t $~=~189.3, 210.9, and 232.2~ms in model R0B0.
The density decreases monotonically with increase in the
radius.  The density gradient is steep around
$ r \, \simeq \, 20~{\rm km} $.  In the outer region
of $ r \, \ga \, 30~{\rm km}$, the density decreases
gently and roughly proportional to $ r ^{-2} $.
Thus we regard the layer of
$\rho \, = \, 10 ^{12} $~g~cm$^{-3}$ as the surface of
PNS for simplicity.  The mass and
radius of PNS are 1.10~M$_\odot$ and
26.6~km, respectively, at $ t $~=~210.9~ms. 
The mass increases by 4.1$\times10^{-2}$~M$_\odot$ between 
$ t $~=~189.3 and 232.2~ms.

Non-radial oscillation has not been excited,
although our numerical code does not assume any symmetry.
This is most likely to be due to short computation time, i.e.,
only 40~ms after the bounce.
Note that the $ \ell $~=~1 mode becomes appreciable 
around 200~ms after the bounce in \citet{burrows06}.

\subsection{Typical Model with Inclined Magnetic Field}

In this subsection we show model R12B12X60 as a typical
example.  The initial rotation period is small compared
to the free-fall timescale,
$ \Omega _0 / \sqrt{4 \pi G \rho _0} \, = \,
1.59 \times 10 ^{-2} $, and initial rotation energy
is much smaller than the gravitational energy
($ |E_{\rm kin}/E_{\rm grav}| \, = \, 5.0 \times 10 ^{-4} $).  The magnetic
energy is also much smaller than the gravitational
energy ($ |E_{\rm mag} /E_{\rm grav}| \, = \, 2.9 \times 10 ^{-4} $).

\begin{figure}
\begin{center}
\includegraphics[width=0.5\textwidth]{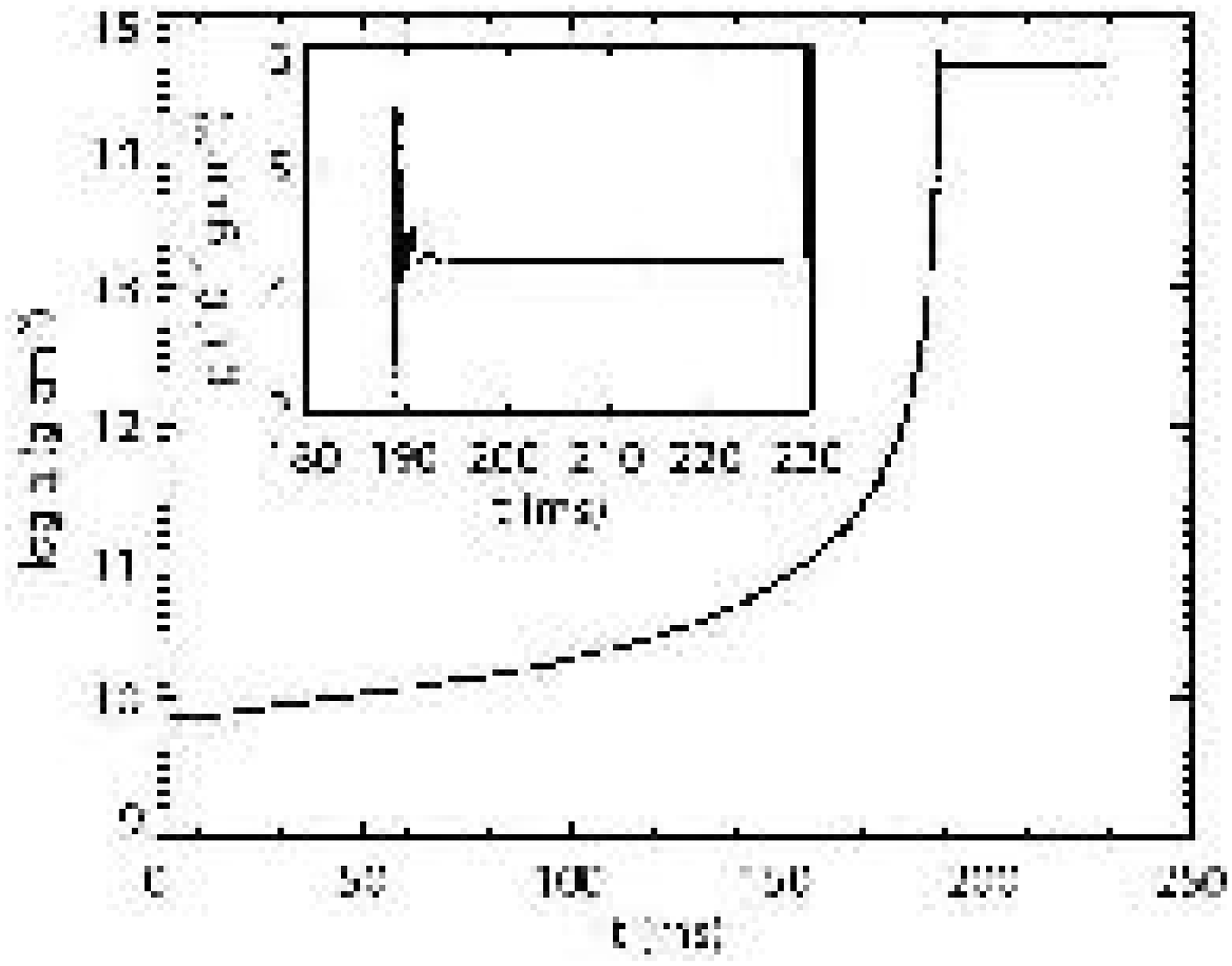}
\end{center}
\caption{The central density is shown as a function of
time for model R12B12X60.\label{rhoc-t}}
\end{figure}

Figure~\ref{rhoc-t} shows the evolution of
central density ($ \rho _c$) for the period of
$ 180~{\rm ms} \, \le \, t \, \le \, 230~{\rm ms} $.
The central density reaches its maximum,
$ 5.49 \times 10 ^{14} $~g~cm$^{-3}$, at
$ t \, = \, 189.05 $~ms as well as in model R0B0.
The period of dynamical collapse is a little longer and
the maximum density is a little lower.
The rotation and magnetic field delays the collapse a little
as has been shown in earlier simulations.
The PNS is only slightly flattened by rotation.

At $ t \, = \, 190.04 $~ms (slightly after the bounce),
the PNS has angular velocity of
$ \Omega _c \, = \, 6.02 \times 10 ^3~{\rm s}^{-1} $ and magnetic field
of $ B _c \, = \, 7.56 \times 10 ^{15}~{\rm G} $ at the center.
The angular velocity and magnetic field increase by a factor of
$ 5.0 \times 10 ^3 $ and $ 3.8 \times 10 ^3 $,
respectively, from the initial values,
while the density increases by a factor of $ 6.1 \times 10 ^4 $.
These enhancements are consistent with the
conservation of the specific angular momentum and
magnetic flux, since the collapse is almost spherical.
The angular velocity and magnetic field increase in proportion
to $ \rho ^{2/3} $ when the collapse is spherical.
The rotation axis changes little.
At this stage the centrifugal force is only 3 \% of the gravity at the
center of the PNS.  The magnetic force is
much weaker than the gravity and than the centrifugal force.

\begin{figure}
\begin{center}
\includegraphics[width=0.88\textwidth]{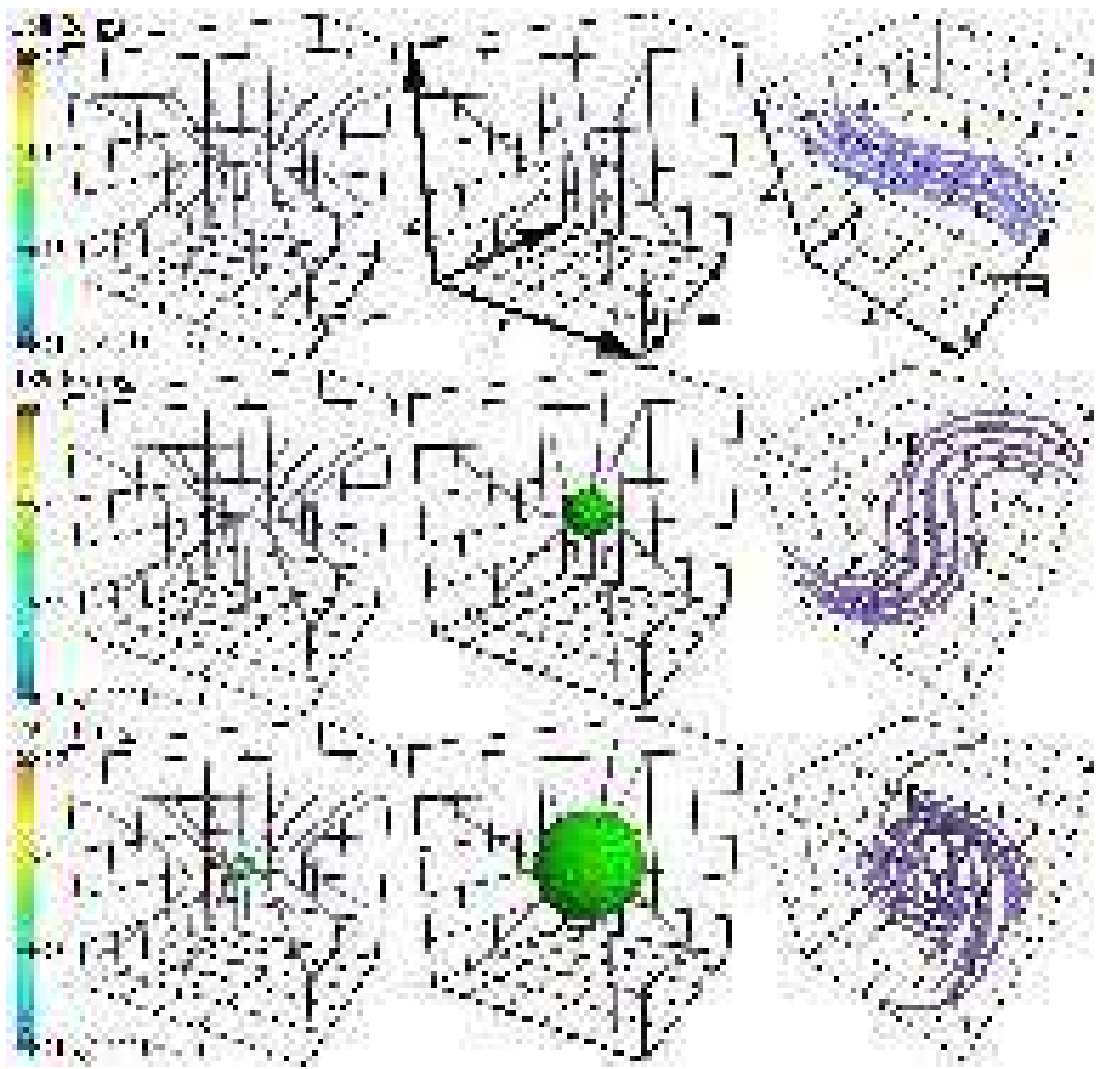}
\end{center}
\caption{The evolution of magnetic field in model R12B12X60.
Each panel denotes the magnetic field lines (purple lines) and
the isovelocity surface of radial velocity, $ v _r $,
at a given stage.  The panels are arranged in the
time sequence from top to bottom.
The top panels denote the stage of $ t $~=~188.28~ms, while
the bottom ones do that of $ t $~=~191.10~ms.
The left panels denote the central cube of (1691~km)$^3$,
while the central and right ones do the zoom-up views of
of (423~km)$^3$ and (26~km)$^3$, respectively.
The left color bars denote the radial velocity on the
surface of the cubes in the unit of 10$^3$~km~s$^{-1}$.
\label{magnetic1}}
\end{figure}

The change in the magnetic field is illustrated in
Figure~\ref{magnetic1}.  The magnetic field is almost radial
near the end of dynamical collapse as shown in the top
panels, in which the purple lines denote the magnetic field lines
at $ t $~=~188.28~ms by bird's eye view.
This is because the magnetic field is stretched in
the radial direction by the dynamical collapse.
The radial component of the magnetic field, $ B _r $, is positive
in the upper half of $ z \, > \, 0 $, while it is negative
in the lower half. Thus the split monopole is a good approximation
to the magnetic field at this stage.

The magnetic field is twisted by the spin of PNS
as shown in the middle and bottom panels of Figure~\ref{magnetic1}.
The azimuthal component of the magnetic field is amplified
to have a large amplitude in the upper atmosphere of the PNS
($ 9~{\rm km} \, \la \, r \,  \la \,  14~{\rm km}$).
The increase in the azimuthal component decreases the plasma beta
down to $ \beta \, \equiv \, P _{\rm gas}/P _{\rm mag} \,
\simeq \, 0.03 $.
The azimuthal component of the magnetic field
is small inside the PNS, since the angular
velocity is nearly constant.  It is also small in
the region very far from the center ($ r \, > \, 60~{\rm km}$)
since the angular velocity is very small.  It is also small
near the rotation axis.  Thus the region of strong twisted
magnetic field has a torus-shape.

\begin{figure}
\includegraphics[width=0.95\textwidth]{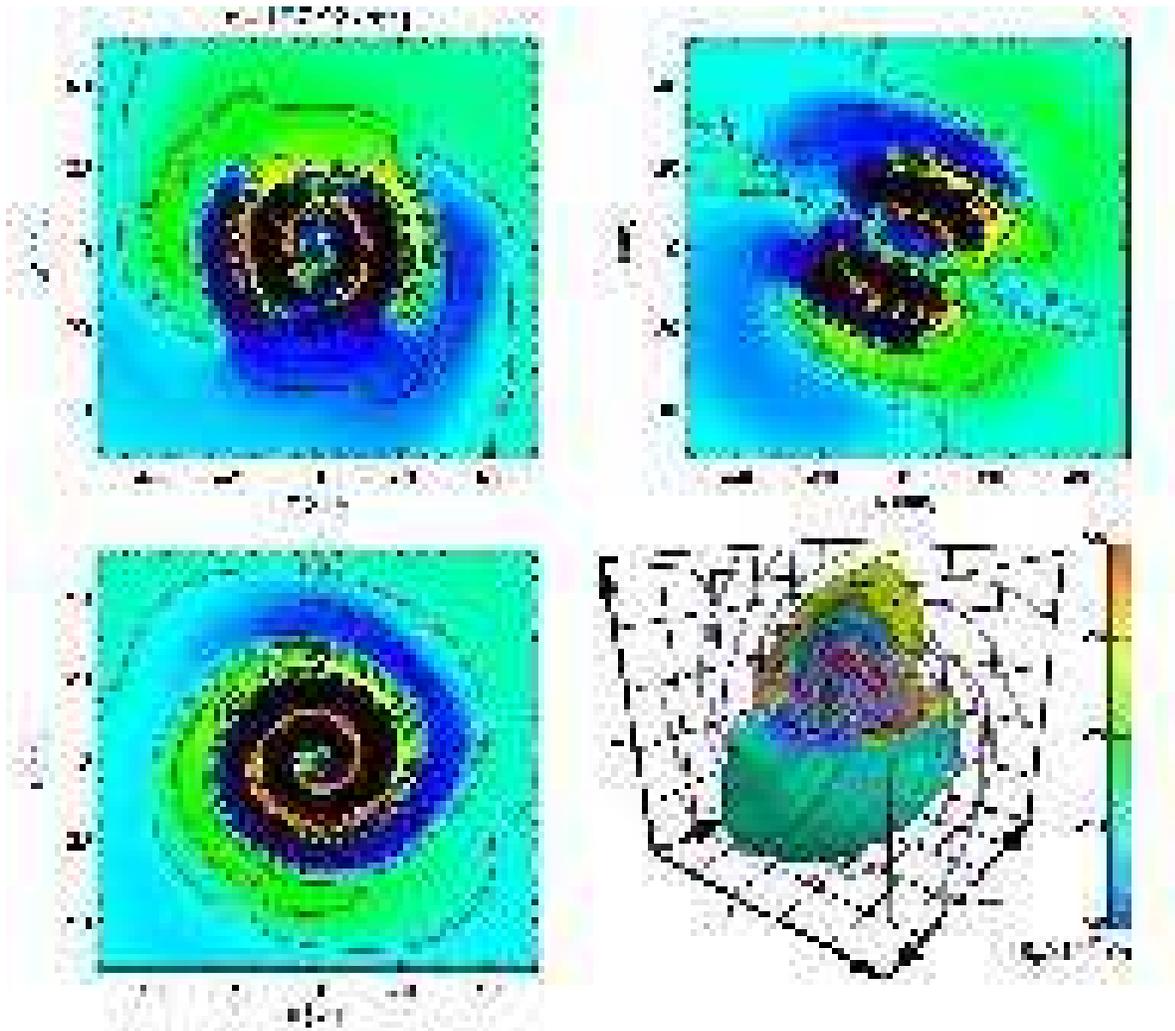}
\caption{Structure of magnetic torus is shown by three cross
sections and a bird's eye view.
The color denotes the azimuthal component of
magnetic field,
$ B _\varphi \, = \, \mbox{\boldmath$e$} _\varphi \cdot
\mbox{\boldmath$B$} $
on the planes of $ x \, = \, 0 $ (upper left),
$ y \, = \, 0 $ (upper right), and $ z \, = \, 0 $
(lower right) at $ t \, = \, 197.92 $~ms.
The contours denote  $B _\varphi$  in unit of $ 10^{15} $~G.
The lower right panels magnetic field lines (purple) and isosurface
of $ B _\varphi $ by the bird's eye view.
\label{bphi-yz}}
\end{figure}

The structure of twisted magnetic field is different
from that in an aligned rotator.  When the initial magnetic
field is aligned to initial rotation axis, the twisted
magnetic field has opposite directions in the upper and
lower halves.  The azimuthal component vanishes on the
equator of rotation and has a large amplitude in the upper and lower tori.
The magnetic field is uni-directional in each torus.
In case of oblique rotator, however, these tori are mixed
into a torus, in which the azimuthal component of magnetic
field is bi-directional as a result.  Figure~\ref{bphi-yz}
shows the distribution of toroidal component of magnetic field,
$ B _\varphi \, = \,  (\mbox{\boldmath$e$} _\varphi
\cdot \mbox{\boldmath$B$}) $,
at $ t \, = \, 197.92~{\rm ms}$, where
$ \mbox{\boldmath$e$} _\varphi \, = \,
\mbox{\boldmath$e$} _\Omega \times \mbox{\boldmath$r$}
/ | \mbox{\boldmath$e$} _\Omega \times \mbox{\boldmath$r$}| $
($ \mbox{\boldmath$e$}_\Omega$ denotes the unit vector
along the initial rotation axis and heads upper left in Figure
\ref{bphi-yz}).
The toroidal component changes its sign alternately with an
average interval of $\sim 5$~km.

\begin{figure}
\centering
\includegraphics[width=0.95\textwidth]{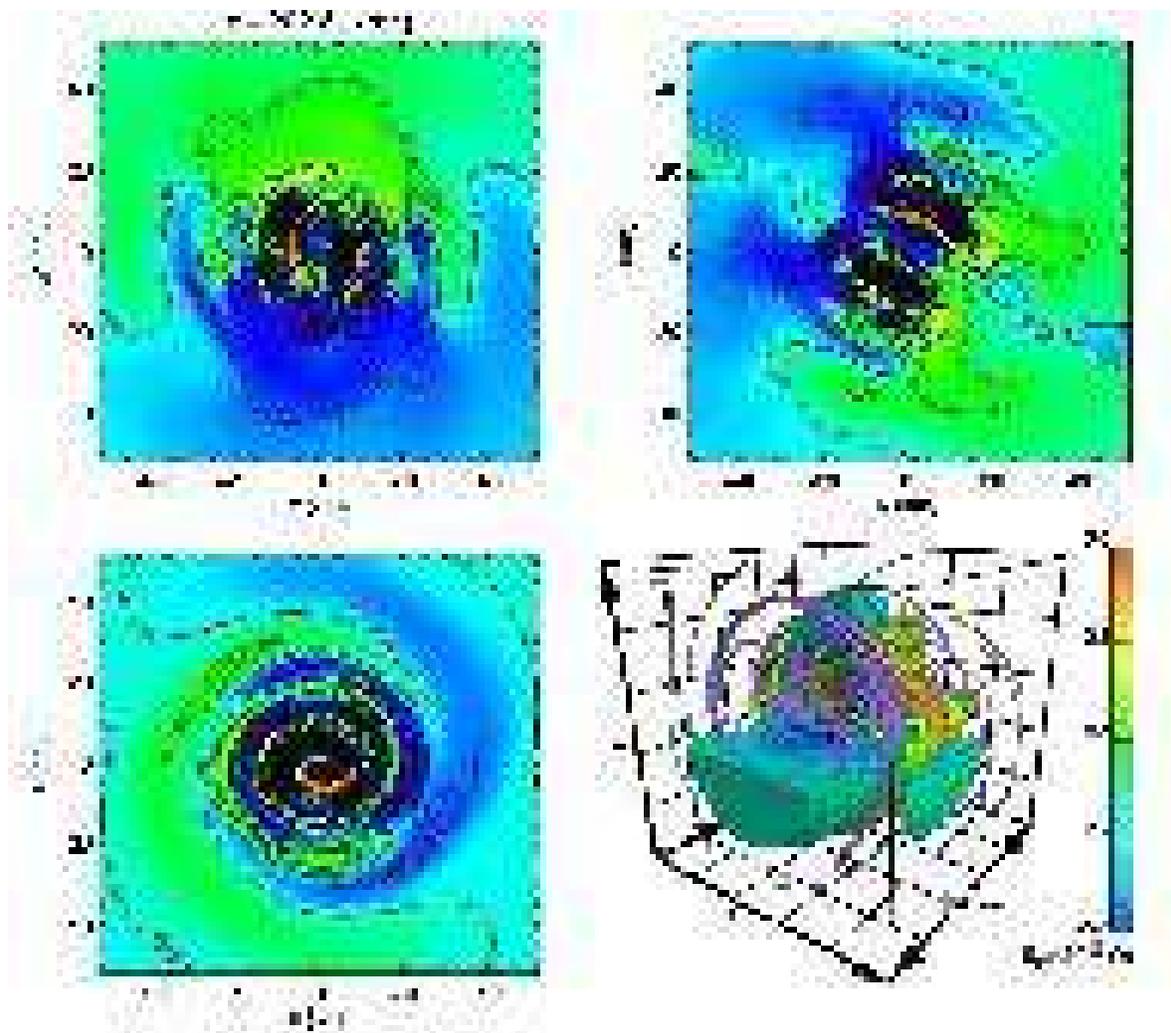}
\caption{The same as Fig.~\ref{bphi-yz} but for the stage
at $ t $~=202.81~ms.\label{bphi-yz2}}
\end{figure}

Figure~\ref{bphi-yz2} is the same Figure~\ref{bphi-yz} but for
the 4.89~ms later stage.  The magnetic field is wound more
tightly inside the PNS and the toroidal component has extended
outward.

A similar magnetic field is obtained semi-analytically
as a model of pulsar magnetosphere by \citet{bogovalov99}.
He approximated the initial magnetic field by split-monopole one
and considered oblique rotation.  As shown in his Figure 4,
the toroidal component changes its sign with a regular interval
in his pulsar wind solution.
His idealized magnetic configuration is realized in our simulation.
The magnetic multi-layers
is inevitably formed when the initial magnetic field is
split-monopole-like and inclined with respect to the rotation
axis.

\begin{figure}
\centering
\includegraphics[width=0.88\textwidth]{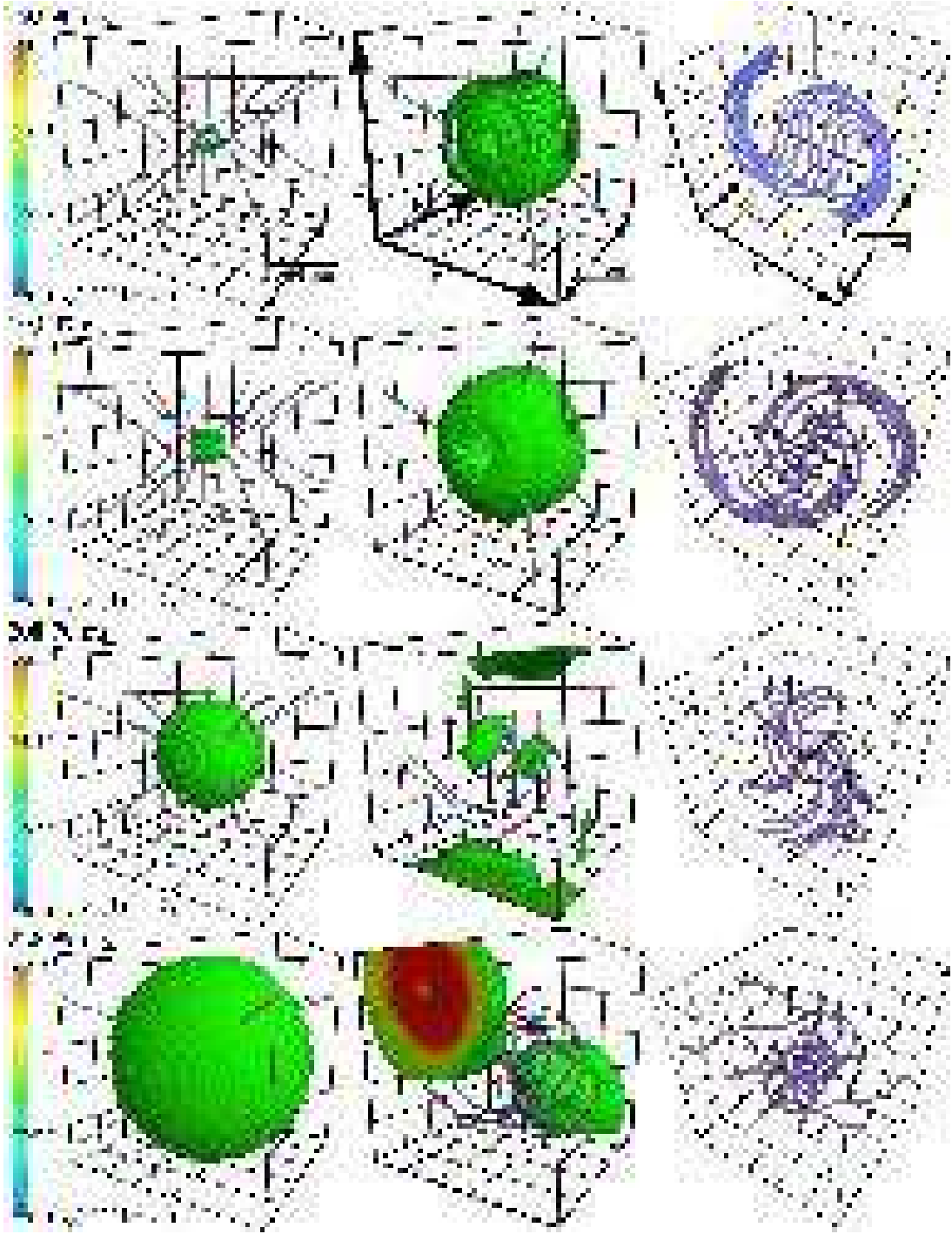}
\caption{The same as Fig.~\ref{magnetic1} but for later stages
of $ t $ = 192.90 ms, 193.97 ms, 208.70 ms, and 228.99 ms.
\label{magnetic2}}
\end{figure}

\begin{figure}
\begin{center}
\includegraphics[width=0.95\textwidth]{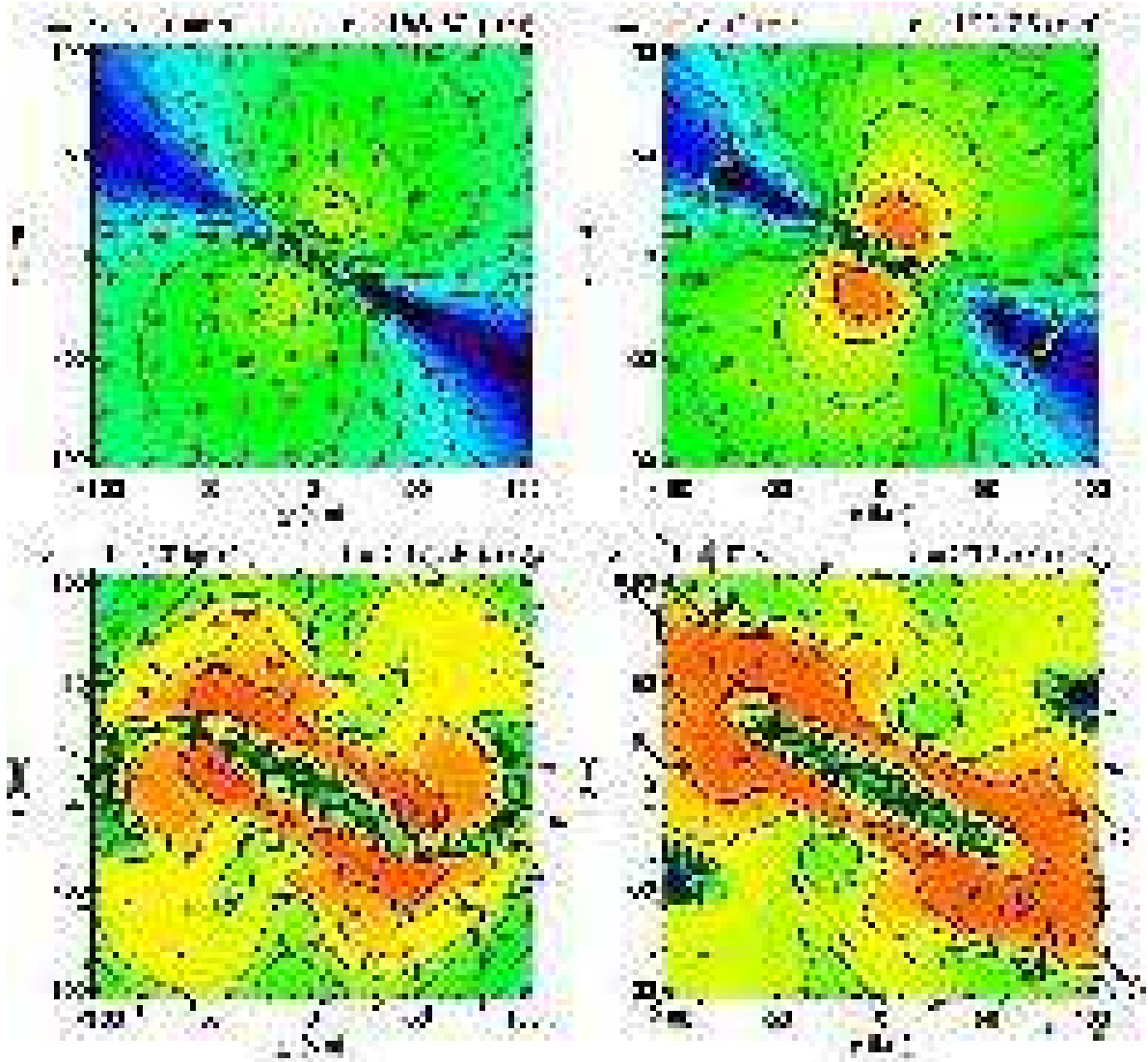}
\end{center}
\caption{The evolution of rotation velocity.  Each panel
denotes the distribution of $ \log \, v _\varphi $~(km~s$^{-1}$)
around the initial rotation axis in the cross section
$ x \, = \, 0$ in model R12B12X60 by color and contours.
The arrows denote the velocity within the plane.
They denote the stages at $ t $~=~188.52, 190.75, 207.56, and 212.96 ms.
\label{vrot-yz}}
\end{figure}

Figure~\ref{magnetic2} shows later evolution of the magnetic
field.  The torus of twisted magnetic field expands slowly and
bipolar jets are launched along the rotation axis.
The jet reaches $ r \, = \, 400~{\rm km}$ at the last stage
of computation ($ t \, = \, 228.99~{\rm ms} $).
The jet velocity exceeds $ 3 \times 10 ^4 $ km s$^{-1}$.
The jets are driven mainly by the magneto-centrifugal
mechanism of \citet{blandford82}.
Figure~\ref{vrot-yz} shows the evolution of rotation
velocity around the initial rotation axis.  The rotation
is nearly rigid in the sphere of $ r \, \la \, 10~{\rm km} $
just before the bounce ($ t $~=~188.52~ms).
The rotation velocity increases up to
$\sim \, 5 \times 10^9 $~km~s$^{-1}$ by the magnetic torque
near the rotation axis.
The magneto-centrifugal force is strong enough to drive jets.
Although the centrifugal force is perpendicular to the
rotation axis, the component perpendicular to the magnetic
field is cancelled by the strong magnetic force.  Thus the
gas is accelerated along the poloidal magnetic field, i.e.,
along the rotation axis by the Blandford-Payne mechanism.

\begin{figure}
\centering
\includegraphics[width=0.95\textwidth]{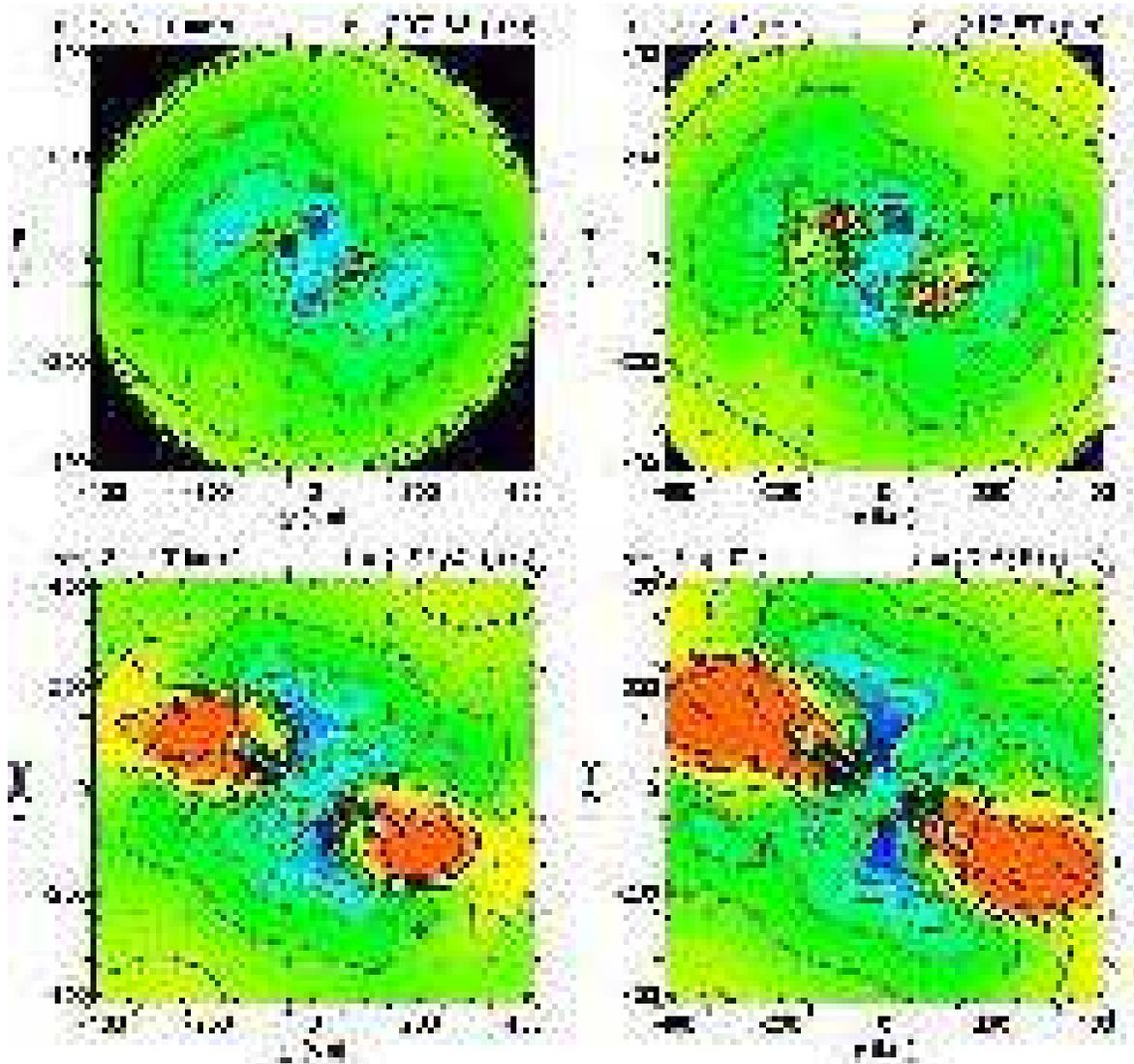}
\caption{The radial velocity ($v _r$)
distribution is denoted by color and contours for
the stages at  $ t $~=~207.56,
212.57, 222.80, and 228.99~ms in model R12B12X60.
The numbers attached to the contours denote
$ v _r $ in unit of $ 10^4 $~km~s$^{-1}$.
The arrows denote the velocity within the cross section.
\label{vrad-yz}}
\end{figure}

\begin{figure}
\begin{center}
\includegraphics[width=0.45\textwidth]
{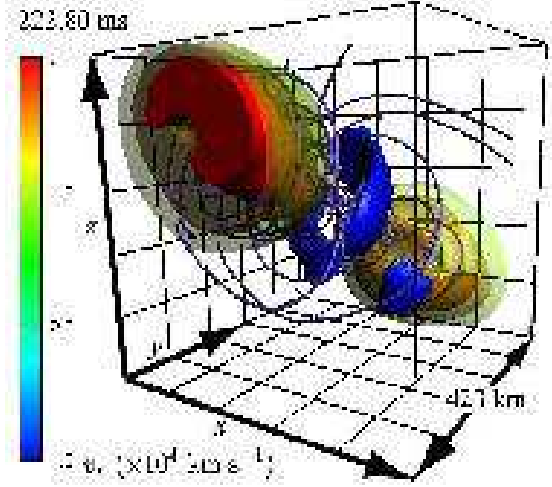}
\end{center}
\caption{Jets and fast radial inflows in model R12B12X60.
The panel denotes the central cube of $ (423~{\rm km}) ^3$
at $ t $~=~222.80~ms by bird's eye view.
The magnetic field lines are denoted by the purple lines,
while the radial velocity is denoted by the isosurfaces.
The blue denotes the fast radial inflow
($v_r \leq -1.0 \times 10^4$ km s$^{-1}$),
while the others do jets ($v_r \geq 2.0 \times 10^4$ km s$^{-1}$).
The color bar is for the radial velocity distribution on the
cube surfaces.\label{jet3d}}
\end{figure}

The increase in the radial velocity follows that in the
rotation velocity.  The acceleration of jets are shown in
Figure~\ref{vrad-yz}.  The radial velocity is still low
at the stage shown in the upper left panel ($ t $~=~207.56~ms).
The high velocity jets emerge not from the PNS surface but from
the upper layer of $ r \, \simeq \, 60$~km.
The mass flux through the sphere of $ r \, = \, 300$~km is 
$\dot{M}$~=~0.0, 7.64, 5.43, 5.58, 3.75, 2.07
M$_\odot$ s$^{-1}$ at
$t$~=~190.75, 207.56, 212.96, 222.80, and 228.99
ms, respectively.
Figure~\ref{jet3d} shows the jets by bird's eye view at the
stage of $ t $~=~222.80~ms.  The jets are bipolar and well collimated.

\begin{figure}
\centering
\includegraphics[width=0.5\textwidth]{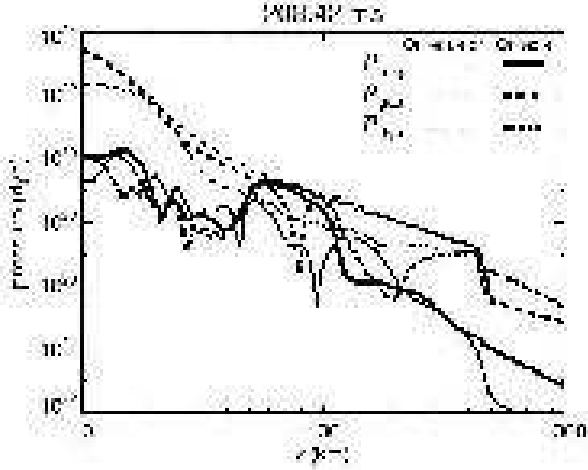}
\caption{Pressure distribution is shown for the stage at
$ t $~=~208.42~ms.  The solid curves denote the magnetic
pressure, $|\mbox{\boldmath$B$}|^2/8\pi$, while the dashed
curves denote the gas pressure.  The dash-dotted curves denote
the dynamical pressure, $ \rho |\mbox{\boldmath$v$}| ^2$.
The thick curves are for the values on the initial rotation
axis while the thin curves for those on equator.\label{pmagpgas}}
\end{figure}

The magnetic force is comparable to the gas pressure
in the jets. Figure~\ref{pmagpgas} denotes the magnetic
pressure distribution at $ t $~=~208.42~ms.
The thick solid curve denotes the magnetic pressure,
$ |\mbox{\boldmath$B$}|^2/8\pi $, along the initial rotation axis,
while thin solid curve does that on the equator.
The magnetic pressure is enhanced by winding in the
range $ 50~{\rm km} \, \la \, r \, \la \, 100~{\rm km} $ on
the axis.  Also the dynamical pressure (rotation energy) is
enhanced in the same region.  Since these energies are
large enough, the jets will extend outwards even if the prompt
shock has stalled around $ r \, \simeq \, 100$~km.

\begin{figure}
\begin{center}
\includegraphics[width=0.5\textwidth]{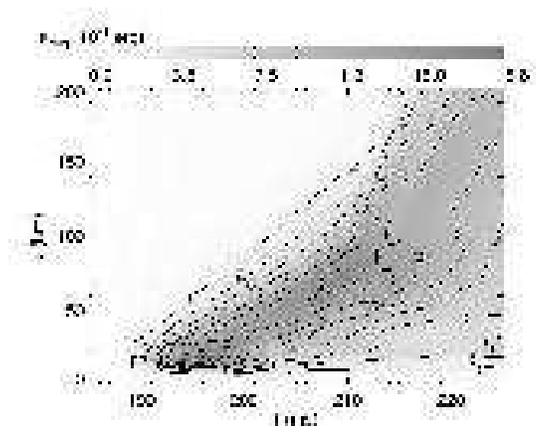}
\end{center}
\caption{The magnetic energy stored in a unit logarithmic
radial distance, $ \varepsilon _{\rm mag} (r,~t) $, is shown by darkness.
The contour levels are set to be
$\Delta \varepsilon _{\rm mag} = 10^{48}$ erg.
The abscissa is the time ($ t $) while the ordinate is radial
distance ($r$).\label{mag-rt-energy}}
\end{figure}

We computed the energy of magnetic field, rotation and
jets to evaluate the efficiency of energy conversion.
The magnetic energy distribution is evaluated by
\begin{equation}
\varepsilon _{\rm mag} (r,~t) \; \equiv \; \int \int r ^3 \,
\frac{|\mbox{\boldmath$B$} \, (r,~\theta,~\varphi,~t)|^2}
{8\pi} \, \sin \theta \, d\theta \, d\varphi \, ,
\end{equation}
i.e., the energy stored in a unit logarithmic radial distance.
The total magnetic energy is expressed as
\begin{equation}
E _{\rm mag} (t) \; = \; \int \,
\varepsilon _{\rm mag} (r,~t) \, d\ln r \, .
\end{equation}
Figure~\ref{mag-rt-energy} shows that the magnetic energy
has a sharp peak of $ \varepsilon _{\rm mag} = 1.88 \times 10 ^{49} $ erg
in the layer of $ r \, \simeq \, 18~{\rm km} $
at $ t~\simeq~196~{\rm ms}$.  The peak of the magnetic
energy shifts outward at the apparent radial velocity of
$  3 \times 10 ^3 $~km~s$^{-1}$, which coincides with
the Alfv\'en velocity.  The peak declines
beyond $ r \, \ga \, 80$ km.

\begin{figure}
\begin{center}
\includegraphics[width=0.5\textwidth]{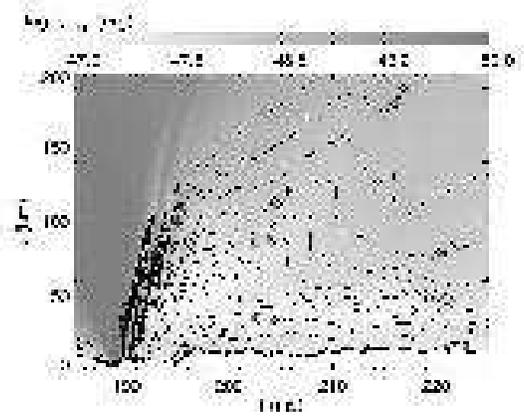}
\end{center}
\caption{The same as Fig.~\ref{mag-rt-energy} but for
the radial kinetic energy, $ \varepsilon _{\rm kin,rad}$.
\label{kinrad-rt-energy}}
\end{figure}

Figure~\ref{kinrad-rt-energy} is the same as
Figure~\ref{mag-rt-energy} but for the radial kinetic energy
stored in a unit logarithmic radial distance,
\begin{equation}
\varepsilon _{{\rm kin,rad}} (r,~t) \; \equiv \;
\int \int r ^3 \, \frac{\rho \, (r,~\theta,~\varphi) \,
| \mbox{\boldmath$n$} \cdot \mbox{\boldmath$v$} \,
(r,~\theta,~\varphi,~t)|^2}
{2} \, \sin \theta \, d\theta \, d\varphi \, ,
\end{equation}
where $ \mbox{\boldmath$n$} \, \equiv \,
\mbox{\boldmath$r$} /|\mbox{\boldmath$r$}| $ denotes
the unit radial vector.  In the region far from the
center, the dynamical collapse dominates in
the radial kinetic energy.
After the bounce ($ t \, > \, 189.72~{\rm ms}$),
the prompt shock propagates and the region of the
dynamical collapse retreats. (The propagation of
prompt shock is mainly due to neglect of neutrino loss.
If we had incorporated the neutrino cooling, the prompt
shock should have stalled around 100-200~km.)
The radial kinetic
energy is small in the region of $ r \, \la \, 60~{\rm km} $
after the bounce.  The bipolar jets increase the radial
kinetic energy in the region of $ r \, \ga \, 80~{\rm km}$.
The rise in $ \varepsilon _{{\rm kin,rad}}$ coincides with the decline
in $ \varepsilon _{\rm mag}$.  This is an evidence that the jets
are accelerated by magnetic force.

\begin{figure}
\centering
\includegraphics[width=0.5\textwidth]{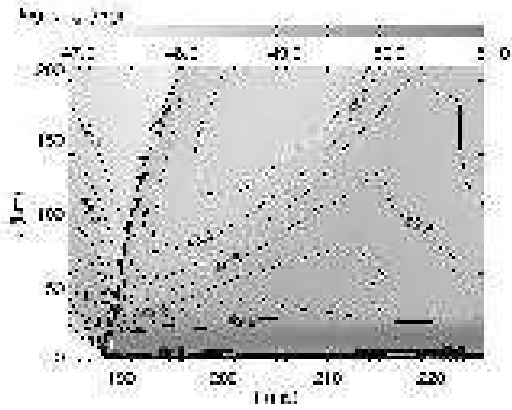}
\caption{The same as Fig.~\ref{mag-rt-energy} but for
the rotational energy, $ \varepsilon _{{\rm kin,rot}}$.
\label{kinrot-rt-energy}}
\end{figure}

Figure~\ref{kinrot-rt-energy} shows the distribution of the
kinetic energy stored in a unit logarithmic radial distance,
\begin{equation}
\varepsilon _{{\rm kin,rot}} (r,~t) \; \equiv \; \int \int r ^3 \,
\frac{\rho \, (r,~\theta,~\varphi) \,
| \mbox{\boldmath$n$} \times \mbox{\boldmath$v$} \,
(r,~\theta,~\varphi,~t)|^2}
{2} \, \sin \theta \, d\theta \, d\varphi \, .
\end{equation}
A large amount of rotation energy is stored in the
region of $ 10~{\rm km} \, \la \, r \, \la \, 20~{\rm km}$.
Only a small fraction of it is converted into the energy
of twisted magnetic field and eventually into the energy of jets.

\begin{figure}
\centering
\includegraphics[width=0.5\textwidth]{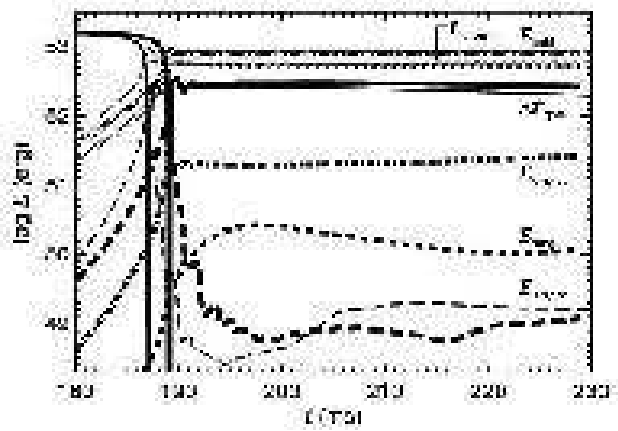}
\caption{The evolution of energy in model R12B12X60.
Each thick curve denotes a component of the energy stored
in the sphere of $ r \, \le \, 63 $~km.  The thin curves
denote those in model R0B0.\label{allenergy}}
\end{figure}

Figure~\ref{allenergy} shows the evolution of energy stored in
the volume of $ r \, \le \, 63~{\rm km} $ for each component.
The gravitational energy, which is evaluated to be
\begin{equation}
E _{\rm grav} \; =  - \; \int _0 ^{63~{\rm km}} \,
\frac{| \mbox{\boldmath$g$} | ^2}{8\pi G} \, dV \, ,
\end{equation}
is the most dominant and the internal energy is comparable.
The thick solid curve denotes, $ \Delta E _{\rm grav} \, \equiv \,
E _{\rm grav} \, - \, E _{\rm grav,\min} $,
the difference from the minimum value, i.e., the
gravitational energy at the stage of the maximum central
density (bounce).
The rotation energy is order of magnitude smaller than them
and the magnetic energy is further smaller.   Only a small
fraction of the rotation energy is converted into magnetic
energy, which is mostly due to toroidal magnetic field.
The energy available
for jet ejection is limited by the conversion factor from
rotation energy to magnetic energy.  The radial kinetic
energy is only of the order of $ \sim 10 ^{49}$~erg in the
period $ t \, \ge \, 195 $~ms since the prompt shock and
jets are outside the region.  For comparison we show the
evolution of the energy stored in model R0B0 by thin curves.

Note that the rotational energy available is much smaller
than those in \citet{obergaulinger06a}.  Since the initial
rotation energy was much larger in their simulation,
the PNS shrunk appreciably after liberating the angular
momentum through magnetic braking.  Even if the rotation
energy were released completely in our model, the PNS
would not shrink appreciably.

\begin{figure}
\centering
\includegraphics[width=0.5\textwidth]
{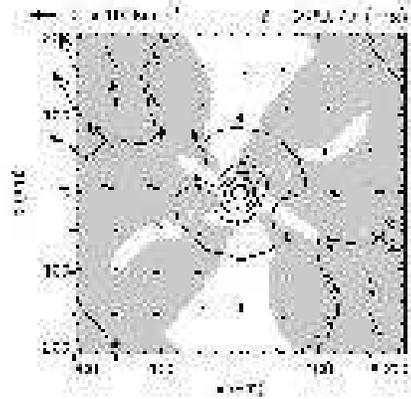}
\caption{The velocity distribution are denoted
by the arrows on the plane of
$ x \, = \, 0$ at $ t \, = \, 229.77~{\rm ms}$. The scale
is shown on the upper left corner.
The darkness denotes the region of positive $ v _r $
while the contours denote isodensity curves and labeled by
$ \log \, \rho $.\label{radial-velocity}}
\end{figure}

When the fast jets are ejected along the initial rotation
axis, two other radial flows are observed.
Figure~\ref{radial-velocity} shows the velocity distribution
on the plane of $ x \, = \, 0 $ at $ t \, = \, 229.77~{\rm ms}$.
One is slow outflow extending near the equator of initial
rotation.  The outflow velocity is approximately
$ 2.5 \times 10 ^4~{\rm km}~{\rm s}^{-1}$.
The other is fast radial inflow located between the jets
and equatorial outflow.  The inflow is less dense and
its dynamical pressure is much smaller than the magnetic
pressure.

As shown earlier, the bipolar jets emanate from
$ r \, \simeq \, 60~{\rm km }$.  In the outer region of
$ r \, \ga \, 60~{\rm km} $, the density is lower than
$ \rho \, \la \, 10 ^{10}~{\rm gm}~{\rm cm}^{-3}$
(see Figure \ref{rho-rt-r0b0} for the average density
distribution in model R0B0) and hence the Alfv\'en velocity is high.
In other words, the magnetic force dominates over the pressure force.
The centrifugal force is also important in the outer region.
If the magnetic field corotates with the PNS,
the rotation velocity is evaluated to be
\begin{equation}
v _\varphi \; = \; 3.6 \times 10 ^4 \,
\left( \frac{\Omega _c}{6 \times 10 ^3~{\rm s}^{-1}}\right)
\, \left( \frac{\varpi}{60~{\rm km}}\right) \, {\rm ~km~s^{-1}},
\end{equation}
where $ \varpi $ denotes the distance from the rotation axis.
The rotation velocity is close to the Keplerian velocity,
\begin{eqnarray}
v _K & = & \sqrt{\frac{GM}{r}} \\
& = & 4.70 \times 10 ^4 \,
\left( \frac{M}{1~{\rm M}_\odot}\right) ^{1/2}
\left( \frac{r}{60~{\rm km}}\right) ^{-1/2} \, {\rm ~km~s^{-1}}
\end{eqnarray}
where $ M $ and $ r $ denote the PNS mass
and the distance from the center, respectively.
In other words, the centrifugal force is comparable
with the gravity.
Thus the foot point of MHD jets coincides with the inner edge of the region
in which the magnetic and centrifugal forces are dominant over the gravity.

As shown in Figure~\ref{bphi-yz} the twisted magnetic field
are ordered and has no structures suggesting development
of magneto-rotational instability \citep[MRI; see e.g.,][]{akiyama03}.
However, this is likely due to limited spatial resolution
and will not exclude the possibility of MRI.
\citet{etienne06} demonstrated that MRI can not grow unless the
cell width is shorter than a tenth wavelength of the fastest growing
mode.  Since the wavelength is evaluated to be
\begin{eqnarray}
\lambda _{\rm MRI} & \approx & 4 \pi \,
\left(r \frac{d\Omega^2}{dr}\right) ^{-1/2} \, \frac{B}{\sqrt{4\pi\rho}} \\
& = &
1.18 \, \left[\frac{d\Omega^2/d\ln r}{\left(3000~{\rm s}^{-1}\right)^2}
\right] ^{-1/2} \,
\left( \frac{B}{10^{15}~{\rm G}} \right) \, \left(
\frac{\rho}{10^{14}~{\rm g}~{\rm cm}^{-3}} \right) ^{-1/2} \,
{\rm km} \, ,
\end{eqnarray}
we need the spatial resolution of $\sim~120$~m to observe MRI.

\subsection{Dependence on $ B _0 $}

\begin{figure}
\centering
\includegraphics[width=0.5\textwidth]{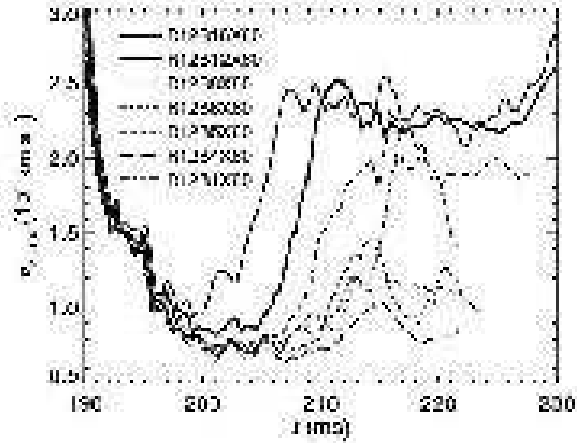}
\caption{
The maximum radial velocity is shown as a function of time for various models
having different initial magnetic field.\label{vr-magnetic}}
\end{figure}

To examine the effect of initial magnetic field
we made 6 models by changing only $ B _0 $ from model
R12B12X60.  Figure~\ref{vr-magnetic} shows
the maximum radial velocity ($ v _{r, \max} $)
as a function of time.
It declines sharply in the period of $t \leq 195$ ms,
since the prompt shock wave slows down.
The early decline is delayed when $B_0$ is larger.
The delay is, however, smaller than 1~ms since the magnetic
energy is much smaller than
the gravitational energy at the PNS formation.
See Table~\ref{bouncetable} for comparison of models at the
bounce.

The late rise in $ v _{r,\max}$, i.e., the launch of
MHD jets depends strongly on $ B _0 $.  When $ B _0 $ is
larger, the radial velocity rises earlier and stays at a
high level.  When $ B _0 $ is smaller than
$ 1.0 \times 10 ^{12}~{\rm G} $, the radial velocity increases
late but decreases soon before reaching a high level.

\begin{figure}
\centering
\includegraphics[width=0.95\textwidth]{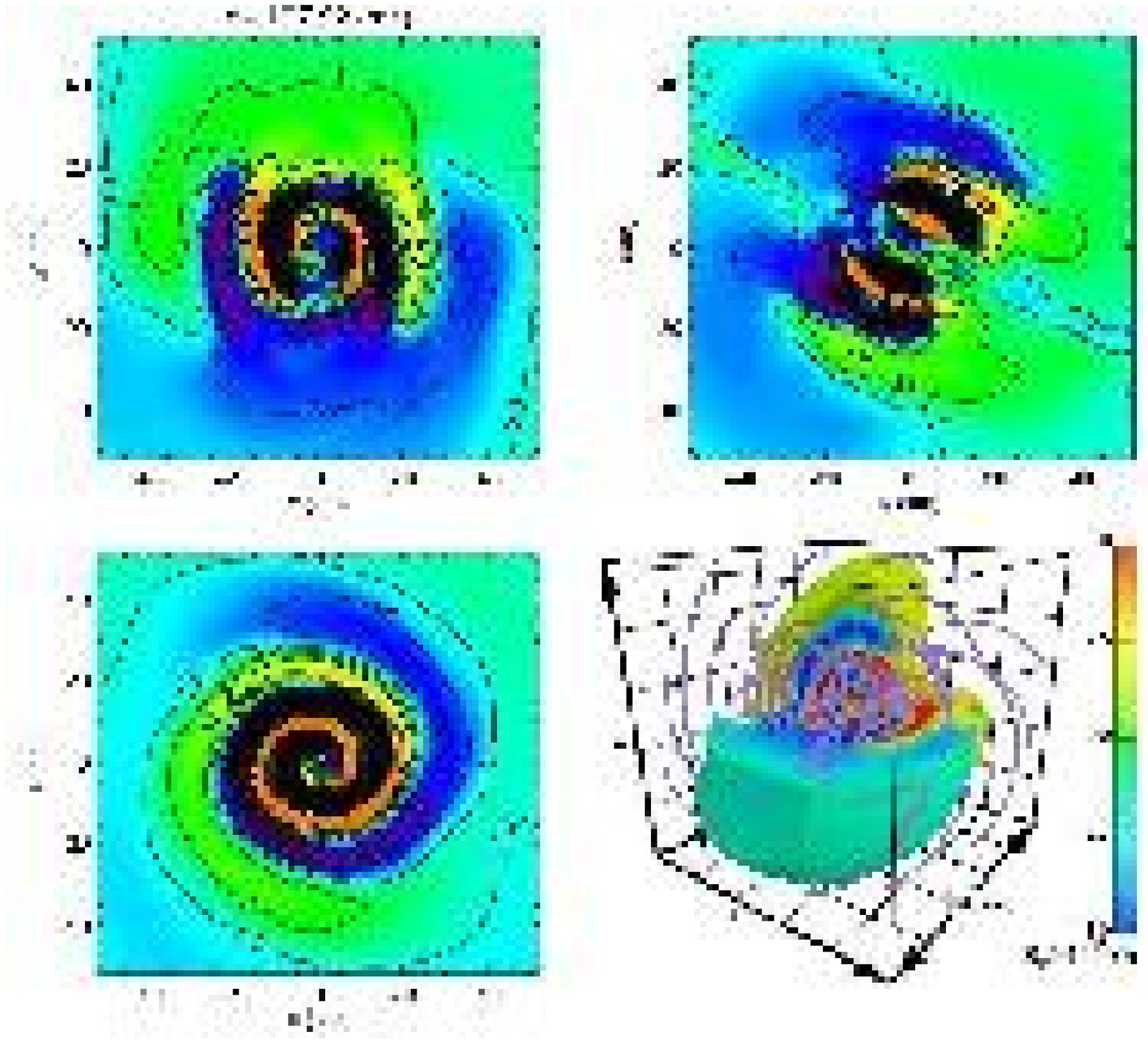}
\caption{The same as Fig.~\ref{bphi-yz} but for
model R12B16X60.\label{bphi-yz-B16}}
\end{figure}

\begin{figure}
\centering
\includegraphics[width=0.5\textwidth]{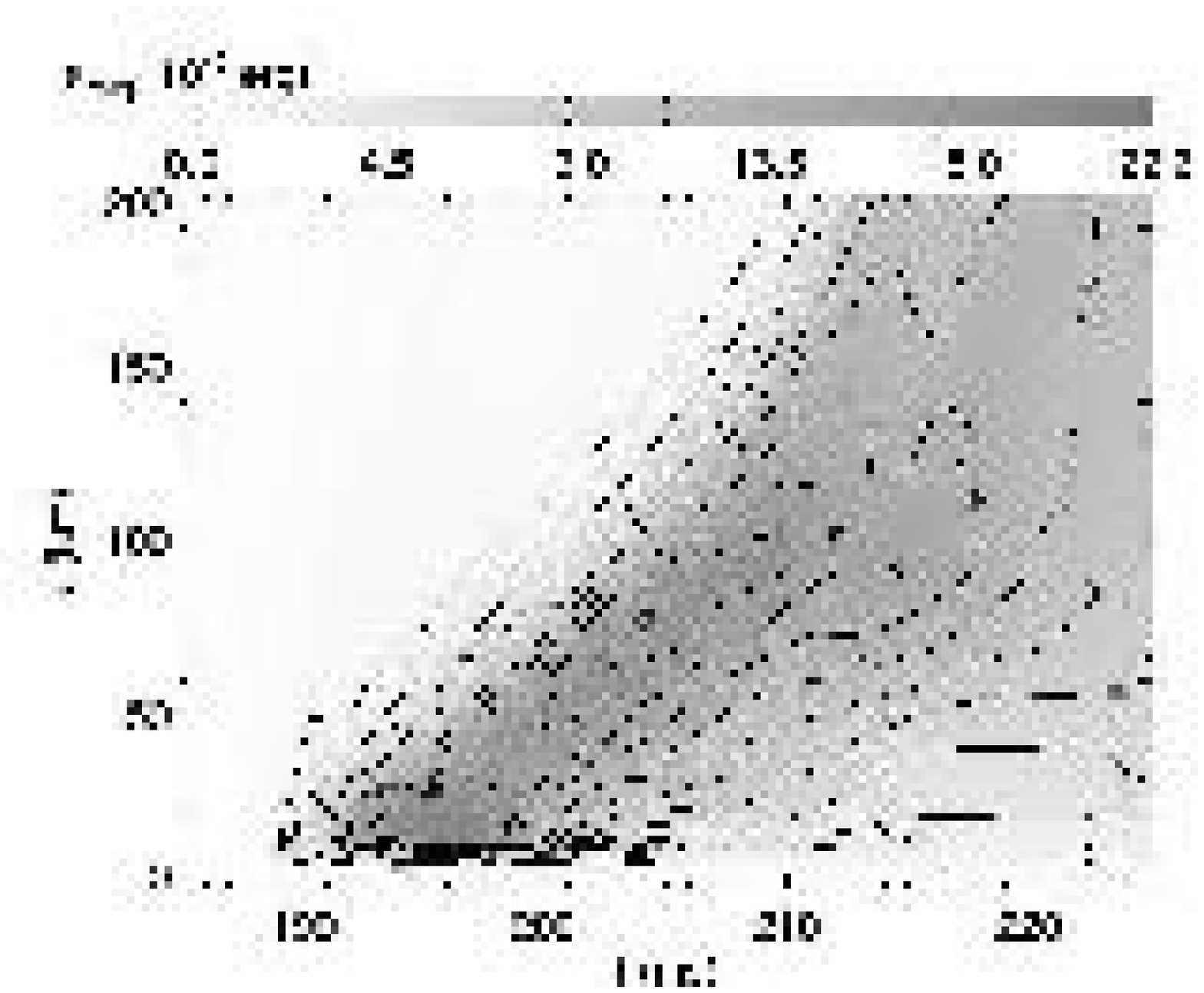}
\caption{The same as Fig.~\ref{mag-rt-energy} but for
model R12B16X60.\label{mag-rt-energy-B16}}
\end{figure}

When $ B _0 $ is large, the magnetic field is less tightly
twisted since the twisted component drifts upward faster.
Figure~\ref{bphi-yz-B16} is the same as Figure~\ref{bphi-yz}
but for model R12B16X60, in which $B_0$ is $4/3$ times larger than
in the standard model R12B12X60.
The toroidal component, $ B _\varphi $,
changes its sign with a longer average interval of
5--6~km on the average,
while it does with an interval of 5~km in the standard model.
At $ t \, \simeq \, 206~{\rm ms}$,
the twisted magnetic field reaches $ r \, = \, 60~{\rm km} $
as shown in Figure~\ref{mag-rt-energy-B16} and the MHD jets initiate.

\begin{figure}
\centering
\includegraphics[width=0.95\textwidth]{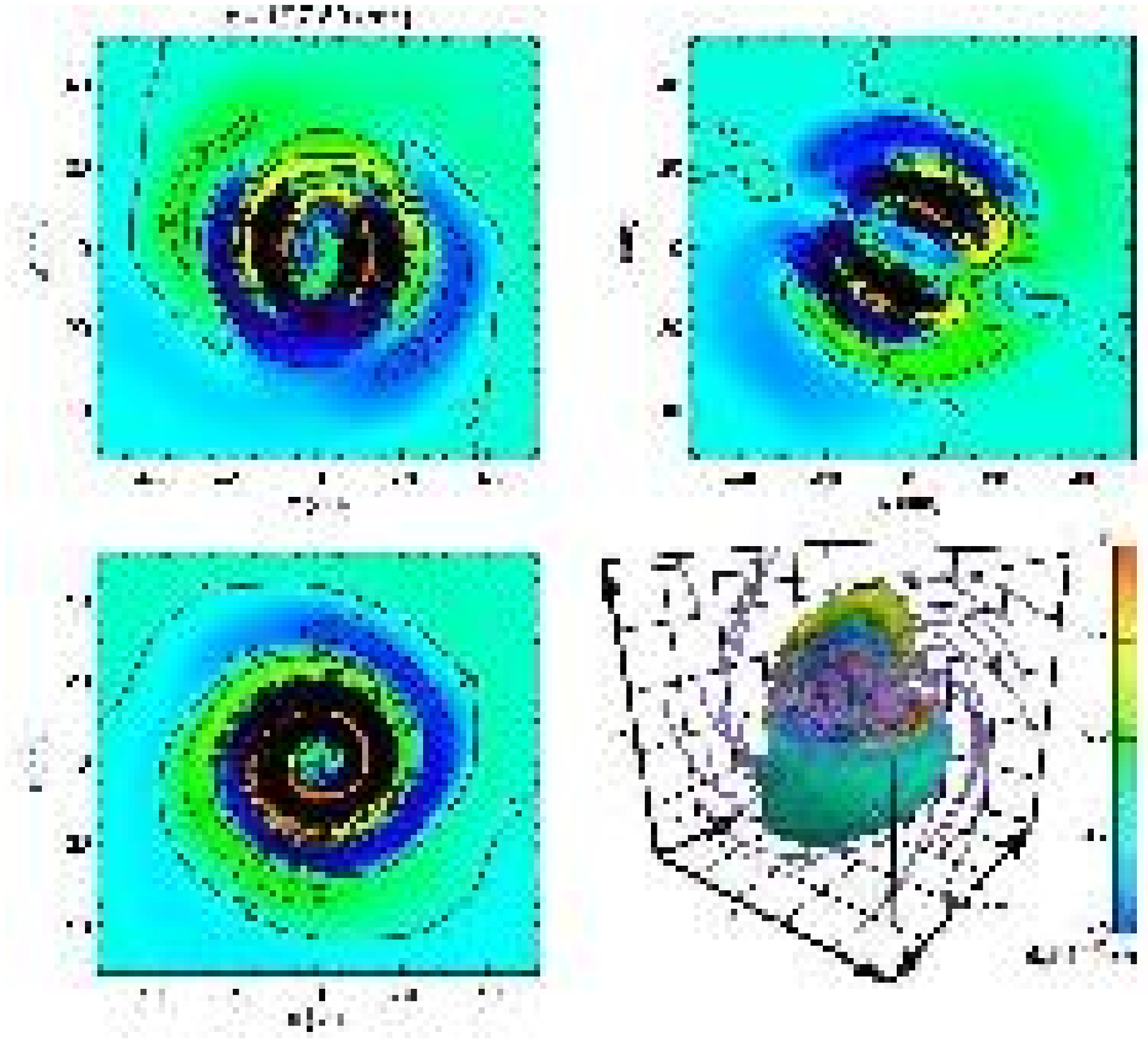}
\caption{The same as Fig.~\ref{bphi-yz} but for
model R12B8X60.\label{bphi-yz-B8}}
\end{figure}

\begin{figure}
\centering
\includegraphics[width=0.5\textwidth]{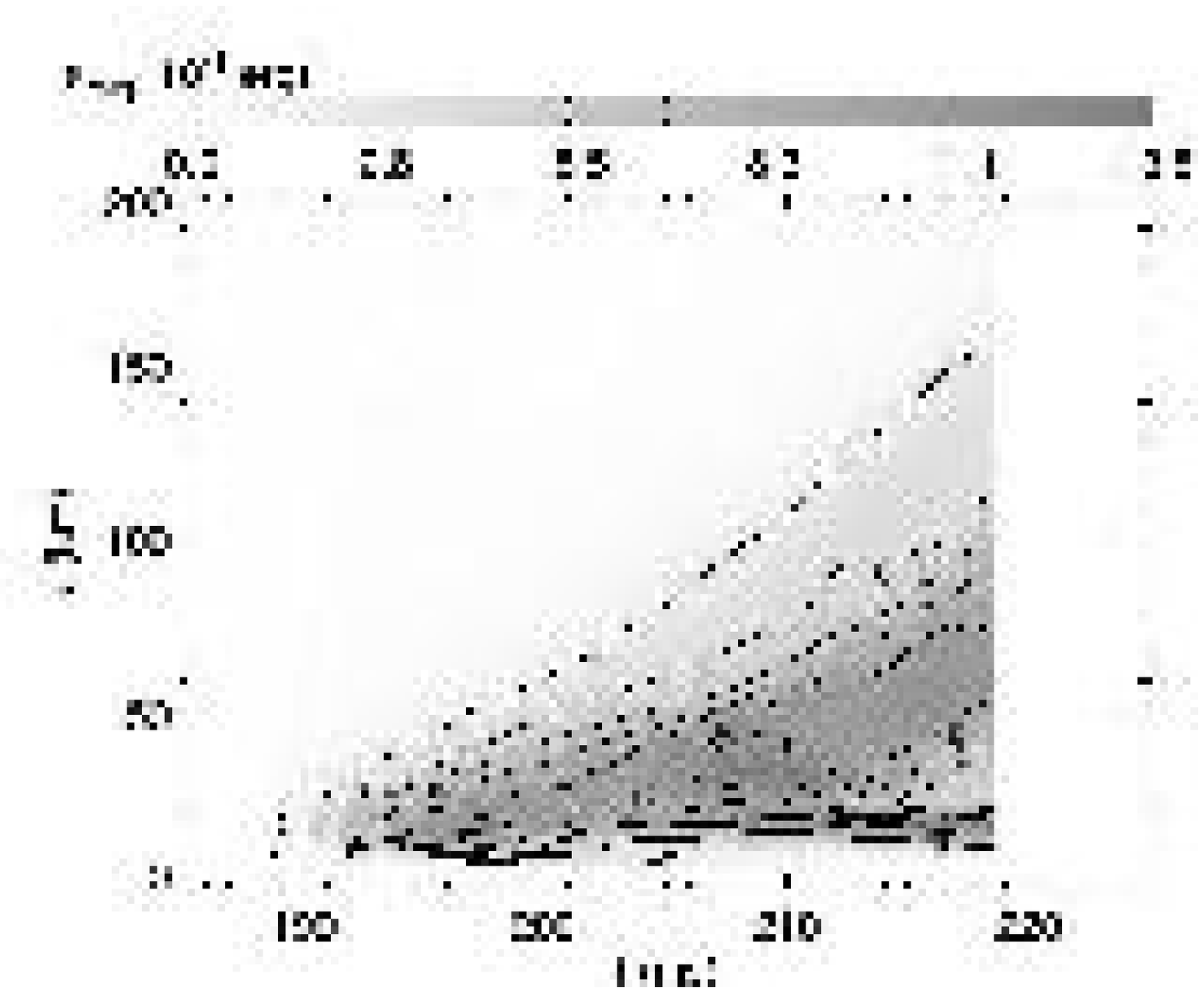}
\caption{The same as Fig.~\ref{mag-rt-energy} but for
model R12B8X60.\label{mag-rt-energy-B8}}
\end{figure}

On the other hand, the magnetic field is more tightly twisted
when $ B _0 $ is smaller.
Figure~\ref{bphi-yz-B8} is the same as Figure~\ref{bphi-yz}
but for model R12B8X60.
The toroidal component, $ B _\varphi $,
changes its sign with a shorter interval of 3--4~km on the average.
The twisted magnetic field rises up slowly
and dwindles as shown in Figure~\ref{mag-rt-energy-B8}.
Accordingly the jets are launched but late and weak
as shown in Figure \ref{bphi-yz-B8}, where the evolution of the maximum
radial velocity, $v_{r,{\rm max}}$, is shown for various models
having different $B_0$.
The weakness of the jet is at least partly due to numerical
diffusion.  Remember that the spatial resolution is 0.826~km
in the central cube of (53.0~km)$^3$.  Thus the numerical
diffusion is appreciably large for the magnetic multi-layers
since the typical interval is less than 10 km.
The MHD jets would be more powerful if the numerical diffusion
were suppressed.

See Table~\ref{finishtable} to compare models at the final
stages.

\subsection{Dependence on $ \Omega _0 $}

To examine the effect of initial rotation,
we made 5 models by changing only $ \Omega _0 $ from model
R12B12X60.  All the models show qualitatively similar
results.  The differences are mainly quantitative.

The initial rotation is twice slower in model R6B12X60 than
in model R12B12X60. Accordingly the PNS has twice
lower angular velocity in model R6B12X60.
The magnetic field is twisted by rotation also in model R6B12X60
but the toroidal component is weaker since the rotation is
slower.  The MHD jets are launched but at a little
later epoch.

\begin{figure}
\centering
\includegraphics[width=0.5\textwidth]{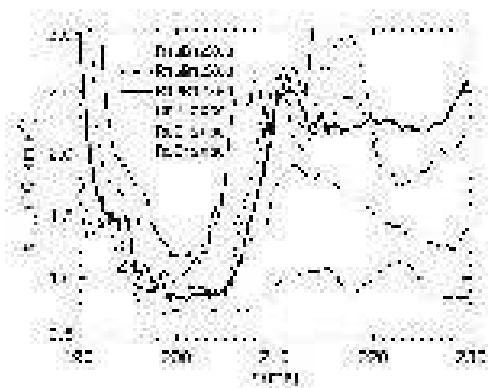}
\caption{The same as Fig.~\ref{vr-magnetic} but
various models having different initial angular velocity.\label{vr-rotation}}
\end{figure}

Figure~\ref{vr-rotation} shows the evolution of the maximum
radial velocity, $ v _{r,\max} $, as a function
of time for various models having different $ \Omega _0 $.
The early decline in $ v _{r,\max} $
 is due to the deceleration of the prompt shock.
The late rise in $  v _{r,\max} $ is due to the launch
of jets.  When $ \Omega _0 $ is larger, the PNS
is formed a little later and the jets are ejected a little
earlier.  The maximum radial velocity is slower when $ \Omega _0 $
is small.

The rotation energy of PNS is proportional
to $ \Omega _0 ^2 $.  It is
$ E _{\rm rot} \, = \, 3.8 \times 10 ^{52}~{\rm erg} $ in
model R18B12X60 while it is
$ E _{\rm rot} \, = \, 4.5 \times 10 ^{51}~{\rm erg} $ in
model R6B12X60.  The energy of the jets, which is evaluated
to be radial kinetic energy of outflowing gas, is also large
in a model having a large $ \Omega _0 $.

\subsection{Dependence on $ \theta _\Omega $}

\begin{figure}[ht]
\centering
\includegraphics[width=0.45\textwidth]
{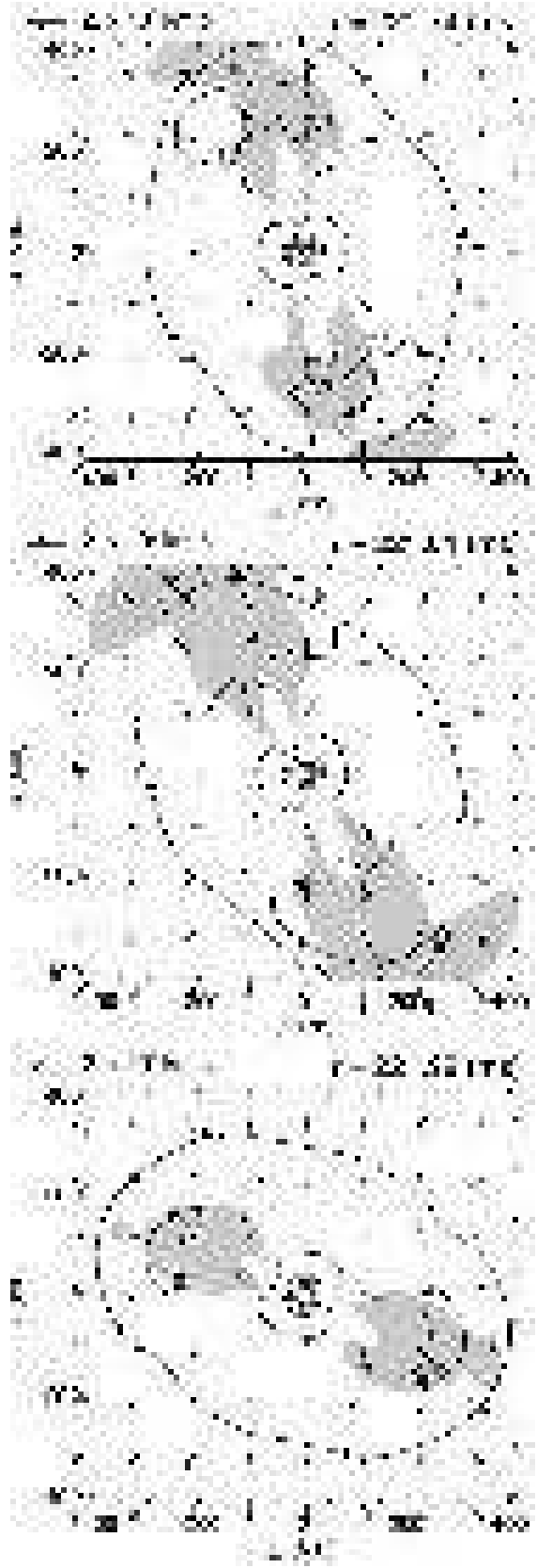}
\caption{Comparison of radial velocity distribution between the
models having different rotation axis ($\theta _\Omega$).
The inclination angle is $\theta _\Omega~=~15^{\circ}$ (top),
30$^{\circ}$ (middle), and
60$^{\circ}$ (bottom), respectively.  The arrows denote the
velocity within the plane and the scale is shown on the top left
of each panel.  The radial velocity exceeds $ v _r \, > \, 5 \times
10 ^3 $~km~s$^{-1}$ in the gray area.
The contours denote the isodensity in the logarithmic
scale and labeled by $\log \, \rho$.\label{jet-inclination}}
\end{figure}

To examine the effect of the initial inclination angle,
we made 5 models by changing only $\theta_\Omega$ from model R12B12X60.
Also in these models, the MHD jets are launched along the initial
rotation axis (see Fig.~\ref{jet-inclination}).

\begin{figure}
\centering
\includegraphics[width=0.95\textwidth]
{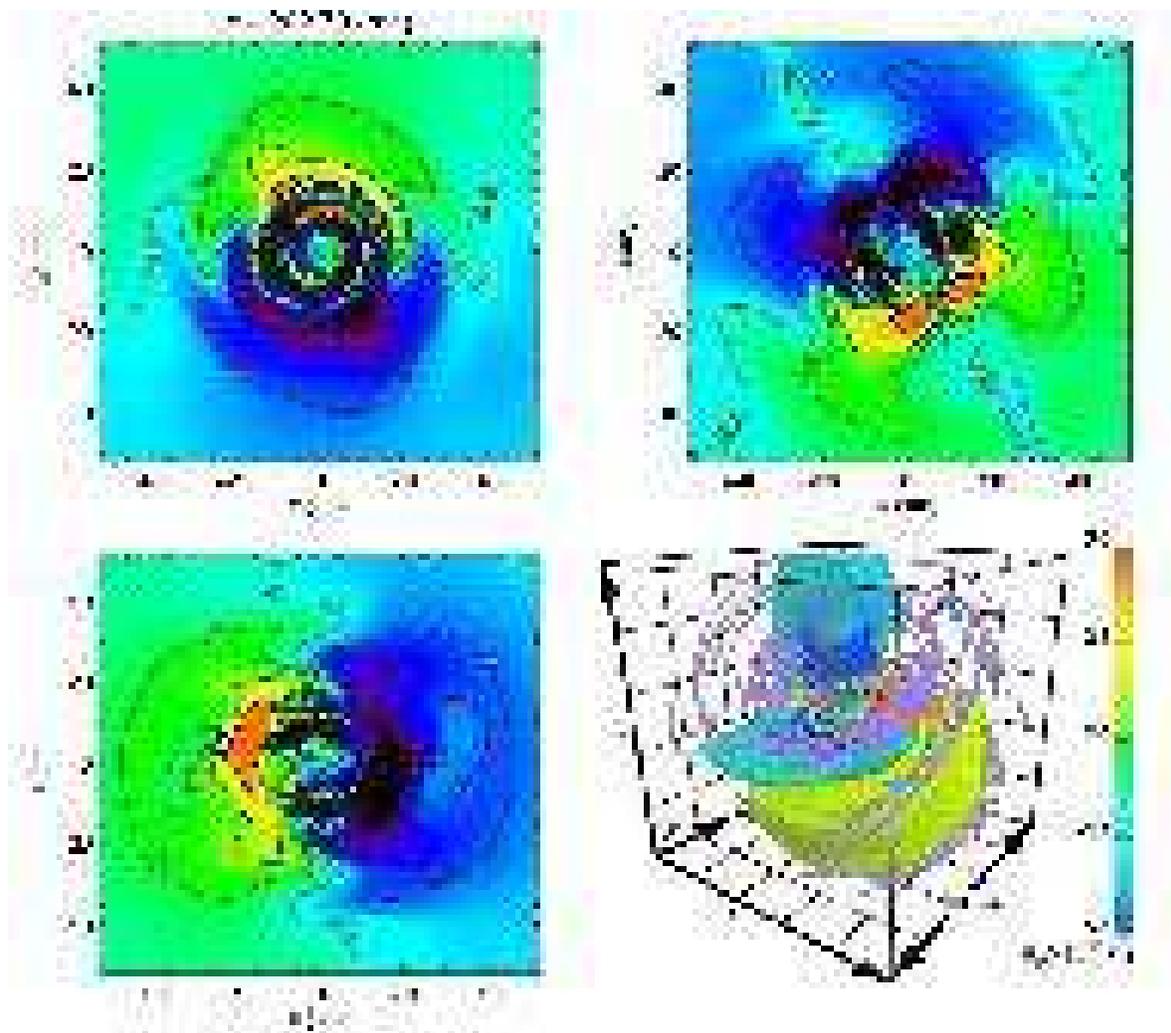}
\caption{The same as Fig.~\ref{bphi-yz} but for model R12B12X30
at $t$~=202.79~ms. \label{bphi-30xi}}
\end{figure}

Figure~\ref{bphi-30xi} denotes the magnetic multi layer formed
in model R12B12X30.  The structure of magnetic multi layer depends
on the inclination angle.  When the inclination angle is smaller,
it is confined in a narrower region around the equator.
The radial interval of changing $ B _\varphi $ depends little
on the inclination angle.

\begin{figure}[ht]
\centering
\includegraphics[width=0.5\textwidth]{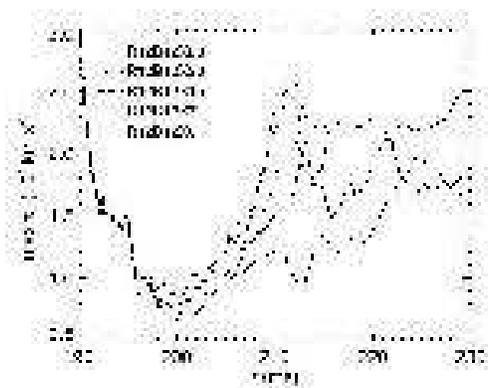}
\caption{The same as Fig.~\ref{vr-rotation}
for various models having different initial inclination angle.
\label{vr-inclination}}
\end{figure}

Figure~\ref{vr-inclination} shows the evolution
of the maximum radial velocity in these models.
When $ \theta _\Omega $ is larger, the maximum radial velocity
rises earlier.  In other words, the MHD jets are launched
earlier when the rotation axis is inclined.

Although the rise is different, the maximum radial velocity
reaches a certain value independently of the inclination
angle.  In other words, the terminal
velocity is independent of the inclination
angle, $ \theta _\Omega $.

\section{SUMMARY AND DISCUSSIONS}

In this paper we have shown three dimensional MHD simulations
of core collapse supernova for the first time.  The numerical
simulations explore the effects of inclined magnetic field
and dependence on the initial magnetic field and rotation.
We summarize the results and discuss the implications.

First we have confirmed that the MHD jets are ejected along
the initial rotation axis.  This is because the energy of
rotation dominates over the magnetic energy at the moment
of PNS formation.  The magnetic force is
too weak to change the rotation axis appreciably.
Thus the magneto-centrifugal force accelerates gas along the
initial rotation axis through the Blandford-Payne mechanism
(see \S 3.2).

If the initial magnetic field were relatively strong, the
rotation axis could change appreciably and also the jet
direction could be different.  A similar problem has been
studied by \citet{machida06} for formation of a protostar.
They considered collapse of a rotating molecular cloud core
having an oblique magnetic field.  They have found that the
evolution of magnetic field and rotation axis depends on their
relative strength.  When the magnetic field is relatively strong,
the magnetic braking acts to align the rotation axis with the
magnetic field.  Then the jets are ejected in the direction
parallel to the initial magnetic field as shown first by
\citet{matsumoto04}.

The relative strength between rotation and magnetic field
can be evaluated from the ratio of the angular velocity to
the magnetic field, $ \Omega / B $.  The ratio remains
nearly constant during the dynamical collapse since the
free-fall timescale is very short.  Both the angular
velocity and magnetic field increase in proportion to the
inverse square of core radius, since both the specific
angular momentum and magnetic flux change little during
the short dynamical collapse phase.
\citet{machida06} proposed $ \Omega _0 / B _0 \, > \,
0.39~G ^{1/2} c _{s} ^{-1} $, as a criterion for the jets parallel
to the initial rotation axis, where $ c _{s} $ denotes the
isothermal sound speed of the molecular cloud.
Since the dynamics of collapse is similar, we can expect
that the criterion holds also for core collapse supernova
if we replace $ c _{s} $ with an appropriate one.
The criterion is rewritten as
\begin{equation}
\frac{\Omega _0 \, c _{s}}{G ^{1/2} \, B _0} \,
> \, 0.39 \, .
\label{criterion}
\end{equation}
The left hand side is evaluated to be
\begin{eqnarray}
\frac{\Omega _0 \, c _{s}}{G ^{1/2} \, B _0} &
= & 3.87 \, \left( \frac{\Omega _0}{1~{\rm s}^{-1}} \right) \,
\left( \frac{B _0}{10^{12}~{\rm G}} \right) ^{-1} \,
\left( \frac{c _{s}}{10^4 \, {\rm km~s}^{-1}}\right) \, .
\end{eqnarray}
The criterion is consistent with our numerical
simulations since the sound speed increases from
$ c _{s} \, = \, 10 ^{4}~{\rm km~s}^{-1} $ to
$ 10 ^{5}~{\rm km~s} ^{-1} $ during the dynamical
collapse.  Although the assumed initial magnetic field is
strong, it is still too weak to change the rotation axis unless
the initial rotation is slow.  Thus it is reasonable that
the rotation is unchanged during the dynamical collapse
since young pulsars are spinning fast.

Next we discuss the fate of magnetic multi-layers
in which the toroidal magnetic field changes
its direction with a regular interval.  The magnetic multi-layers
are a natural outcome of oblique rotation
as shown in the previous section.  These layers are potentially
unstable against reconnection, although no features are seen for
reconnection.  It is also interesting to study the
magnetic multi-layers with a higher spatial resolution.
If the magnetic fields are reconnected, a large amount of the
magnetic energy is released to lead an explosive process.

Next we discuss the lag between the bounce and jet ejection.
When the initial magnetic field is weaker, the lag is longer
as already shown in \citet{ardeljan04}, \citet{moiseenko06},
 and \citet{burrows07}.
Our numerical simulations
have suggested that the lag is related to the Alfv\'en transit time;
the MHD jets are ejected when the twisted magnetic field reaches
a certain radius, i.e., 60~km in our simulations.
When the initial magnetic field is weak, the Alfv\'en transit
time is longer and the magnetic field is amplified for a longer
duration before the jet ejection.
Since the larger amplification compensates for a weak seed
field, almost the same amount of toroidal magnetic field is
generated irrespectively of the initial magnetic field
strength.  This implies that strong MHD jets can be
ejected even if the initial magnetic field is weak.
If we could suppress numerical diffusion by improving
resolution, strong MHD jets should be ejected also in
model R12B5X60 and others having a weaker initial
magnetic field as discussed in the previous section.

The Alfv\'en transit time is evaluated to be
\begin{eqnarray}
\tau _{A} & \equiv & \int ^{r _j} \frac{1}{v _A} \, dr \\
& = & \int ^{r_j} \frac{\sqrt{4\pi \rho}}{B _r} \, dr \, ,
\end{eqnarray}
where $ r _j~$($\simeq$ 60 km) denotes the radius at the foot point of
the jets.  The radial magnetic field decreases in proportion to
the inverse square of the radius ($ \propto \, r ^{-2}$)
since the split monopole is a good approximation for
the magnetic field at the epoch of PNS
formation.  The density decreases also roughly
proportional to the inverse square of the radius
($ \propto \, r ^{-2}$) in the upper atmosphere of
the PNS (see Figure \ref{rho-rt-r0b0}).
Accordingly the Alfv\'en velocity is roughly proportional to
the radius, $ v _{A} \, \propto \, r $ and the Alfv\'en
transit time depends only weakly on $ r _j $.

The above argument suggests a factor controlling the jet
energy.  The rotation energy of the PNS is
much larger than the energy of jets.  If the magnetic field
is twisted for a longer period, i.e., if the Alfv\'en
transit time is longer, a larger fraction of the
rotation energy is converted into the jet energy.
The Alfv\'en transit time can be extended if the magnetic
field is twisted in a deep interior of the PNS.

\acknowledgments

We would like to thank M. Shibata and S. Yamada for their
valuable comments on our numerical works.  We also thank
R. Matsumoto, S. Miyaji, and members of the astrophysical
laboratory of Chiba University for stimulating discussion
and encouragement.
This work is supported financially in part by the Grant-in-Aid
for the priority area from the Ministry of Education, Culture,
Sports, Science and Technology of Japan (17030002).
HM acknowledges Chiba University for the financial support
for his attendance to the Meeting on Astrophysics of Compact
Objects at Huangshan in July 2007, in which the main result of
this paper was presented orally.

\appendix

\section{An Approximate Riemann Solution of the MHD Equations for
Non-ideal Equation of State}

For the approximate EOS of \citet{takahara82}, we
have found a numerical flux which satisfies the
property U of \citet{roe81}.  The numerical flux is the
same as that of \citet{cargo97} except for the
correction term,
\begin{equation}
G \; = \; \overline{\varepsilon _{\rm cold}} \, + \,
\overline{\rho} \, \frac{\Delta \varepsilon _{\rm cold}}{\Delta \rho}
\, - \, \frac{1}{\gamma _{\rm t} \, -  \, 1} \,
\frac{\Delta P _{\rm cold}}{\Delta \rho} \,
\end{equation}
where
\begin{equation}
\varepsilon _{\rm cold} \; = \; \int _0 ^\rho
\frac{P _{\rm cold}}{\rho ^2} \, d\rho \, .
\end{equation}
Here the bared symbols denote the Roe average while the
symbols with the capital delta denote the differences
between the two adjacent cells.  The average specific
enthalpy, $ \bar{H} $, should be replaced with
$ \bar{H} \, - \, G $ in the computation of $ a $ and
$ a _*$.  Accordingly, the characteristic speeds propagating in the
$ x $-direction are evaluated to be
\begin{equation}
a ^2 \; = \; (\gamma _t \, - \, 1) \, \left(\bar{H} \, - \, G
\, - \, \frac{\bar{u} ^2 \, + \, \bar{v} ^2 \, + \, \bar{w}^2}{2}
\, + \, \frac{B _x ^2 \, + \, \bar{B} _y ^2 \, + \, \bar{B} _z ^2}
{4 \pi \bar{\rho}} \, - \, \delta{b} ^2
\right) \, ,
\end{equation}
\begin{eqnarray}
a _* ^2 & = & (\gamma _t \, - \, 1) \, \left(\bar{H} \, - \, G
\, - \, \frac{\bar{u} ^2 \, + \, \bar{v} ^2 \, + \, \bar{w}^2}{2}
\, - \, \delta{b} ^2 \right) \nonumber \\
& \; & \; - \, (\gamma _t \, - \, 2) \, \left(
\frac{B _x ^2 \, + \, \bar{B} _y ^2 \, + \, \bar{B} _z ^2}
{4\pi \bar{\rho}}
\right) \, ,
\end{eqnarray}
where
\begin{equation}
\delta b ^2 \; = \; \frac{\gamma _t \, - \, 1}{\gamma _t \, - \, 2}
\, \frac{\Delta B _y ^2 \, + \, \Delta B _z ^2}{4 \pi \,
(\sqrt{\rho _{\rm L}} \, + \, \sqrt{\rho _{\rm R}}) ^2} \, .
\end{equation}
Here, $ (u, \, v, \, w) $ and $ (B _x, \, B _y, \, B _z)$ denote
the velocity and magnetic field, respectively.  The
symbols, $ \Delta B _y $ and $ \Delta B _z $, denote the
difference between the two adjacent cells, while the
symbols, $ \rho _{\rm L} $ and $ \rho _{\rm R} $ denote the
density in the left hand side and that in the right hand side, respectively.

At the same time, the correction term $ G $
should be added to the last component of the
eigen vector of the entropy wave.  This correction term
is similar to that obtained by \citet{nobuta99} for the
numerical flux of hydrodynamical equations.
Thus the right eigen vector for the entropy wave is expressed as,
\begin{equation}
^{t}\mbox{\boldmath$r$} _{\rm entropy} \; = \;
\left(1, \, \bar{u}, \, \bar{v}, \bar{w}, \, 0, \, 0, \, 0, \,
\frac{\bar{u} ^2 \, + \, \bar{v} ^2 \, + \, \bar{w} ^2}{2} \,
+ \, G \right) \, .
\end{equation}

\clearpage

\clearpage

\begin{table}[ht]
\caption{Model parameters for $ P _c$. See Eq. \ref{eos3}.}
\begin{center}
\begin{tabular}{cccc}
\hline
$ i $ & $ \rho _i $ & $ K _i $ & $\gamma _i $\\
\hline
1 & $ 4.0 \times 10 ^9 $ & 
$ 1.767 \times 10 ^{27} $ & $4/3$\\
2 & $ 1.0 \times 10 ^{12} $ & 
$ 2.446 \times 10 ^{30} $ & 1.31 \\
3 & $ 2.8 \times 10^{14} $ & 
$ 4.481 \times 10 ^{33} $ & $4/3$ \\
4 & $ 1.0 \times 10^{15} $ & 
$ 1.080 \times 10 ^{35} $ & 2.5 \\
\hline
\end{tabular}
\end{center}
\label{eos-table}
\end{table}

\begin{deluxetable}{lllllllll}
\label{modeltable}
\tablecolumns{8}
\tablewidth{0pc}
\tablecaption{Summary of the models.  The initial gravitational
energy is $ W \, = \, -4.75 \times 10^{51}$~erg in all the 
models.  The models underlined are shown in detail in the main text.}
\tablehead{
model  & $\Omega_0$     & $E_{\rm kin}$ & $\left| E_{\rm kin}/E_{\rm grav}\right|$ &$B_0$&$E_{\rm mag}$& $\left|E_{\rm mag}/E_{\rm grav}\right|$ & $\theta_{\Omega}$ \\
& (s$^{-1}$) &(erg)& (\%)                & (G) &(erg)&   (\%)               &
}
\startdata
\underline{R0B0}
          &  0   &  0  &  0  &  0  &  0  &  0  &  \nodata \\
\underline{R12B12X60}
          & 1.2  & 2.2$ \times 10^{48}$ & $5.0 \times 10^{-4}$ & 2.0$ \times 10^{12}$ & 1.4$ \times 10^{48}$ & $2.9 \times 10^{-4}$ & 60$^{\circ}$ \\
R18B12X60 & 1.8  & 4.9$ \times 10^{48}$ & $1.1 \times 10^{-3}$ & 2.0$ \times 10^{12}$ & 1.4$ \times 10^{48}$ & $2.9 \times 10^{-4}$ & 60$^{\circ}$ \\
R15B12X60 & 1.8  & 3.4$ \times 10^{48}$ & $1.1 \times 10^{-3}$ & 2.0$ \times 10^{12}$ & 1.4$ \times 10^{48}$ & $2.9 \times 10^{-4}$ & 60$^{\circ}$ \\
R8B12X60  & 0.81 & 9.6$ \times 10^{47}$ & $2.0 \times 10^{-4}$ & 2.0$ \times 10^{12}$ & 1.4$ \times 10^{48}$ & $2.9 \times 10^{-4}$ & 60$^{\circ}$ \\
R6B12X60  & 0.61 & 5.4$ \times 10^{47}$ & $1.1 \times 10^{-5}$ & 2.0$ \times 10^{12}$ & 1.4$ \times 10^{48}$ & $2.9 \times 10^{-4}$ & 60$^{\circ}$ \\
R3B12X60  & 0.31 & 1.4$ \times 10^{47}$ & $2.0 \times 10^{-4}$ & 2.0$ \times 10^{12}$ & 1.4$ \times 10^{48}$ & $2.9 \times 10^{-4}$ & 60$^{\circ}$ \\
\underline{R12B16X60}
          & 1.2  & 2.2$ \times 10^{48}$ & $5.0 \times 10^{-4}$ & 2.7$ \times 10^{12}$ & 2.5$ \times 10^{48}$ & $5.8 \times 10^{-4}$ & 60$^{\circ}$ \\
\underline{R12B8X60}
          & 1.2  & 2.2$ \times 10^{48}$ & $5.0 \times 10^{-4}$ & 1.3$ \times 10^{12}$ & 6.2$ \times 10^{47}$ & $1.4 \times 10^{-4}$ & 60$^{\circ}$ \\
R12B6X60  & 1.2  & 2.2$ \times 10^{48}$ & $5.0 \times 10^{-4}$ & 1.0$ \times 10^{12}$ & 3.5$ \times 10^{47}$ & $8.1 \times 10^{-5}$ & 60$^{\circ}$ \\
R12B5X60  & 1.2  & 2.2$ \times 10^{48}$ & $5.0 \times 10^{-4}$ & 8.4$ \times 10^{11}$ & 2.4$ \times 10^{47}$ & $5.6 \times 10^{-5}$ & 60$^{\circ}$ \\
R12B4X60  & 1.2  & 2.2$ \times 10^{48}$ & $5.0 \times 10^{-4}$ & 6.7$ \times 10^{11}$ & 1.6$ \times 10^{47}$ & $3.6 \times 10^{-5}$ & 60$^{\circ}$ \\
R12B1X60  & 1.2  & 2.2$ \times 10^{48}$ & $5.0 \times 10^{-4}$ & 1.7$ \times 10^{11}$ & 9.7$ \times 10^{45}$ & $2.3 \times 10^{-6}$ & 60$^{\circ}$ \\
R12B12X30 & 1.2  & 2.2$ \times 10^{48}$ & $5.0 \times 10^{-4}$ & 2.0$ \times 10^{12}$ & 1.4$ \times 10^{48}$ & $2.9 \times 10^{-4}$ & 30$^{\circ}$ \\
R12B12X15 & 1.2  & 2.2$ \times 10^{48}$ & $5.0 \times 10^{-4}$ & 2.0$ \times 10^{12}$ & 1.4$ \times 10^{48}$ & $2.9 \times 10^{-4}$ & 15$^{\circ}$ \\
R12B12X7  & 1.2  & 2.2$ \times 10^{48}$ & $5.0 \times 10^{-4}$ & 2.0$ \times 10^{12}$ & 1.4$ \times 10^{48}$ & $2.9 \times 10^{-4}$ & 7$^{\circ}$ \\
R12B12X0  & 1.2  & 2.2$ \times 10^{48}$ & $5.0 \times 10^{-4}$ & 2.0$ \times 10^{12}$ & 1.4$ \times 10^{48}$ & $2.9 \times 10^{-4}$ & 0$^{\circ}$ \\
\enddata
\end{deluxetable}

\begin{deluxetable}{lllllllll}
\label{bouncetable}
\tablecolumns{8}
\tablewidth{0pc}
\tablecaption{Comparison of models at the bounce.}
\tablehead{
model
  & $t$
  & $T/\left|W\right|$
  & ${\cal M}/\left|W\right|$
  & $T$
  & ${\cal M}$
  & $\omega_{\rm max}$
  & $B_{\rm max}$
 \\
 & (ms)
 & (\%)
 & (\%)
 & ($10^{50}$ erg)
 & ($10^{50}$ erg)
 & (kHz)
 & ($10^{15}$ G)
}
\startdata
R0B0 & $191.9$ & $0.5854$ & 0.000 & $8.359$ & 0.000 & $0.1421$ & 0.000 \\
R12B12X60 & $192.2$ & $2.137$ & $0.1002$ & $30.46$ & $1.428$ & $29.97$ & $17.96$ \\
R18B12X60 & $192.2$ & $4.315$ & $0.07094$ & $57.21$ & $0.9405$ & $38.21$ & $13.48$ \\
R15B12X60 & $194.1$ & $2.703$ & $0.1679$ & $37.72$ & $2.344$ & $38.44$ & $21.25$ \\
R8B12X60 & $191.9$ & $1.309$ & $0.07185$ & $18.61$ & $1.021$ & $26.50$ & $16.08$ \\
R6B12X60 & $193.1$ & $0.9306$ & $0.07122$ & $13.24$ & $1.013$ & $16.72$ & $18.46$ \\
R3B12X60 & $193.2$ & $0.6340$ & $0.04088$ & $9.019$ & $0.5815$ & $6.329$ & $18.04$ \\
R12B16X60 & $192.7$ & $2.033$ & $0.1721$ & $28.91$ & $2.447$ & $28.20$ & $23.09$ \\
R12B8X60 & $192.3$ & $2.150$ & $0.05520$ & $30.53$ & $0.7840$ & $28.39$ & $13.41$ \\
R12B6X60 & $192.6$ & $2.022$ & $0.1000$ & $28.17$ & $1.394$ & $7.656$ & $16.68$ \\
R12B5X60 & $192.3$ & $2.180$ & $0.02334$ & $30.96$ & $0.3314$ & $27.82$ & $9.320$ \\
R12B4X60 & $220.7$ & $2.887$ & $0.01407$ & $41.06$ & $0.2001$ & $31.03$ & $9.416$ \\
R12B1X60 & $192.5$ & $2.127$ & $0.08790$ & $30.23$ & $1.249$ & $35.31$ & $16.50$ \\
R12B12X30 & $194.9$ & $1.986$ & $0.1384$ & $27.98$ & $1.950$ & $42.41$ & $16.26$ \\
R12B12X15 & $194.9$ & $1.981$ & $0.1423$ & $27.91$ & $2.005$ & $41.17$ & $16.83$ \\
R12B12X7 & $192.7$ & $2.036$ & $0.03528$ & $28.22$ & $0.4889$ & $43.35$ & $6.467$ \\
\enddata
\end{deluxetable}

\begin{deluxetable}{llllllllll}
\label{finishtable}
\tablecolumns{9}
\tablewidth{0pc}
\tablecaption{Comparison of models at the final stage.}
\tablehead{
model
  & $t$
  & $T/\left|W\right|$
  & ${\cal M}/\left|W\right|$
  & $T$
  & ${\cal M}$
  & $\Delta E_{\rm grav}$
  & $\omega_{\rm max}$
  & $B_{\rm max}$
 \\
 & (ms)
 & (\%)
 & (\%)
 & ($10^{50}$ erg)
 & ($10^{50}$ erg)
 & ($10^{50}$ erg)
 & (kHz)
 & ($10^{15}$ G)
}
\startdata
R0B0 & $232.2$ & $0.1552$ & 0.000 & $2.312$ & 0.000 & $61.53$ & $5.123$ & 0.000 \\
R12B12X60 & $229.0$ & $2.277$ & $0.1835$ & $32.44$ & $2.615$ & $1.002$ & $27.80$ & $12.80$ \\
R18B12X60 & $215.8$ & $3.497$ & $0.1955$ & $48.38$ & $2.705$ & $57.54$ & $32.85$ & $15.89$ \\
R15B12X60 & $224.4$ & $2.988$ & $0.2084$ & $42.11$ & $2.937$ & $13.59$ & $32.59$ & $14.21$ \\
R8B12X60 & $228.7$ & $1.200$ & $0.1547$ & $17.33$ & $2.235$ & $23.50$ & $19.85$ & $13.99$ \\
R6B12X60 & $231.9$ & $0.7440$ & $0.1066$ & $10.94$ & $1.568$ & $47.14$ & $12.32$ & $14.89$ \\
R3B12X60 & $233.2$ & $0.2592$ & $0.06470$ & $3.853$ & $0.9616$ & $63.81$ & $5.220$ & $11.56$ \\
R12B16X60 & $230.5$ & $2.010$ & $0.2382$ & $28.77$ & $3.409$ & $8.921$ & $24.99$ & $12.74$ \\
R12B8X60 & $227.7$ & $2.514$ & $0.1548$ & $35.67$ & $2.197$ & $1.261$ & $38.35$ & $13.37$ \\
R12B6X60 & $224.9$ & $1.773$ & $0.1817$ & $23.26$ & $2.383$ & $81.68$ & $12.02$ & $12.99$ \\
R12B5X60 & $220.5$ & $2.550$ & $0.1770$ & $35.96$ & $2.495$ & $9.983$ & $29.23$ & $17.35$ \\
R12B4X60 & $223.7$ & $3.011$ & $0.01702$ & $43.08$ & $0.2435$ & $7.955$ & $31.27$ & $11.05$ \\
R12B1X60 & $217.6$ & $1.933$ & $0.2232$ & $27.24$ & $3.145$ & $12.26$ & $38.46$ & $15.85$ \\
R12B12X30 & $230.6$ & $2.208$ & $0.3719$ & $30.72$ & $5.174$ & $17.36$ & $43.41$ & $14.15$ \\
R12B12X15 & $209.0$ & $1.712$ & $0.3392$ & $23.83$ & $4.722$ & $16.61$ & $34.29$ & $28.97$ \\
R12B12X7 & $231.8$ & $1.864$ & $0.2876$ & $23.79$ & $3.669$ & $110.0$ & $37.51$ & $34.89$ \\
\enddata
\end{deluxetable}

\end{document}